\pdfoutput=1
\documentclass[twocolumn,twocolappendix,floatfix]{aastex63}
\usepackage{CJKutf8}

\usepackage{amsmath,amsthm,amssymb}
\usepackage{graphicx}
\usepackage{svg}
\usepackage{xcolor}
\usepackage{microtype}
\usepackage{needspace}
\usepackage{hyperref}
\nonstopmode

\newcommand{\uat}[2]{\href{http://astrothesaurus.org/uat/#2}{#1 (#2)}}

\newcommand{\teff}{$\rm{T_{eff}}$}

\newcommand{\logg}{log $g$}
\newcommand{\feh}{[Fe/H]}
\newcommand{\mgfe}{[Mg/Fe]}

\begin{document}
\begin{CJK*}{UTF8}{gbsn}

\title{Disentangling Stellar Age Estimates from Galactic Chemodynamical Evolution}

\correspondingauthor{Jeff Shen}
\email{shenjeff@princeton.edu}

\author[0000-0001-6662-7306]{Jeff Shen}
\affiliation{David A. Dunlap Department of Astronomy \& Astrophysics, University of Toronto, 50 St. George Street, Toronto, ON M5S 3H4, Canada}
\affiliation{Department of Statistical Sciences, University of Toronto, 9th Floor, Ontario Power Building, 700 University Ave, Toronto, ON M5G 1Z5, Canada}
\affiliation{Department of Astrophysical Sciences, Princeton University, 4 Ivy Lane, Princeton, NJ 08544, USA}

\author[0000-0003-2573-9832]{Joshua S. Speagle (\begin{CJK*}{UTF8}{gbsn}沈佳士\ignorespacesafterend\end{CJK*})}
\affiliation{Department of Statistical Sciences, University of Toronto, 9th Floor, Ontario Power Building, 700 University Ave, Toronto, ON M5G 1Z5, Canada}
\affiliation{David A. Dunlap Department of Astronomy \& Astrophysics, University of Toronto, 50 St. George Street, Toronto, ON M5S 3H4, Canada}
\affiliation{Dunlap Institute for Astronomy \& Astrophysics, University of Toronto, 50 St. George Street, Toronto, ON M5S 3H4, Canada}
\affiliation{Data Sciences Institute, University of Toronto, 17th Floor, Ontario Power Building, 700 University Ave, Toronto, ON M5G 1Z5, Canada}

% \author[0000-0002-6411-8695]{Neige Frankel}
% \affiliation{David A. Dunlap Department of Astronomy \& Astrophysics, University of Toronto, 50 St. George Street, Toronto, ON M5S 3H4, Canada}
% \affiliation{Canadian Institute for Theoretical Astrophysics, University of Toronto, 60 St. George Street, Toronto, ON M5S 3H8, Canada}
% \affiliation{Max Planck Institute for Astronomy, K{\"o}nigstuhl 17, D-69117 Heidelberg, Germany}

\author[0000-0001-8108-0935]{J. Ted Mackereth}
\affiliation{David A. Dunlap Department of Astronomy \& Astrophysics, University of Toronto, 50 St. George Street, Toronto, ON M5S 3H4, Canada}
\affiliation{Dunlap Institute for Astronomy \& Astrophysics, University of Toronto, 50 St. George Street, Toronto, ON M5S 3H4, Canada}
\affiliation{Canadian Institute for Theoretical Astrophysics, University of Toronto, 60 St. George Street, Toronto, ON M5S 3H8, Canada}

\author[0000-0001-5082-9536]{Yuan-Sen Ting (\begin{CJK*}{UTF8}{gbsn}丁源森\ignorespacesafterend\end{CJK*})}
\affiliation{Research School of Astronomy \& Astrophysics, Australian National University, Canberra, ACT 2611, Australia}
\affiliation{School of Computing, Australian National University, Canberra, ACT 2601, Australia}

\author[0000-0001-6855-442X]{Jo Bovy}
\affiliation{David A. Dunlap Department of Astronomy \& Astrophysics, University of Toronto, 50 St. George Street, Toronto, ON M5S 3H4, Canada}
\affiliation{Dunlap Institute for Astronomy \& Astrophysics, University of Toronto, 50 St. George Street, Toronto, ON M5S 3H4, Canada}

\begin{abstract}
  Stellar ages are key for determining the formation history of the Milky Way, but are difficult to measure precisely.
  Furthermore, methods that use chemical abundances to infer ages may entangle the intrinsic evolution of stars with the chemodynamical evolution of the Galaxy.
  In this paper, we present a framework for making probabilistic predictions of stellar ages, and then quantify the contribution of both stellar evolution and Galactic chemical evolution to those predictions using SHAP values.
  We apply this interpretable prediction framework to both a simulated Milky Way sample containing stars in a variety of evolutionary stages and an APOGEE-mocked sample of red clump stars.
  We find that in the former case, stellar evolution is the dominant driver for age estimates, while in the latter case, the more restricted evolutionary information causes the model to proxy ages through the chemical evolution model.
  We show that as a result of the use of non-intrinsic Galactic chemical information, trends estimated with the predicted ages, such as the age-metallicity relation, can deviate from the truth.
\end{abstract}


\shorttitle{Interpretable Stellar Ages}
\shortauthors{Shen et al.}

\needspace{5\baselineskip}
\section{Introduction}\label{sec:intro} 

Measuring the chronochemodynamics of stars allows us to reconstruct a timeline of events in the Galaxy.
Of particular importance are stellar ages; good measurements of stellar ages are necessary to understand stellar evolution and the formation history of the Milky Way, and to identify new stellar populations within the Galaxy \citep{Mackereth2019, Lu2021, Lu2021a, Ness2019, Aguirre2018}.
However, stellar ages are particularly difficult to measure, as they are not directly related to any observable (see \citealt{Soderblom2010} for a review).
Instead, we can only measure properties that are correlated with ages.
For example, ages can be obtained using gyrochronology \citep[e.g.,][]{Barnes2003, Angus2019} uses the empirical relation between a star's rotation period and its age, and asteroseismology \citep[e.g.,][]{Lebreton2008,Aguirre2018}, which uses measurements of oscillation frequencies to probe the properties of the interiors of stars, which are related to age.

With the advent of surveys like APOGEE \citep{Majewski2017}, measurements of many stellar surface properties are available, and they can be used to infer ages.
Surface properties like \logg\ and \teff\ describe stellar type, so they can be used in conjunction with theoretical stellar evolution models to get information about the age of a star \citep{Bressan2012, DeglInnocenti2016, Choi2016}.
Stellar evolution also changes surface abundances through physical processes like dredge-up, whereby CNO products (nitrogen-rich and carbon-poor) are mixed into the atmosphere of a star through convection.
The CNO process is also mass dependent, and as a result, abundance ratios like [C/N] can give insight into stellar evolution and be used to estimate ages \citep{Salaris2015}.
Thus, information about stellar evolution (i.e., age) is also correlated with \feh, [$\alpha$/Fe], and other chemical abundances \citep{Nissen2017, Casali2019}.
However, measured stellar abundances also depend on the material that a star was formed from; chemical properties of the interstellar medium vary across time and location within the Galaxy \citep{Spina2016, Mena2019, Ness2019}.
For example, \cite{Nissen2017} find for \textit{Kepler} LEGACY stars in the Galactic disk that [Mg/Fe] is correlated with age because the deposition of iron into the ISM from SNIa is delayed relative to the production of Mg from SNII.
Thus, ages that are inferred from certain chemical abundances are entangled with Galactic chemical evolution.

For understanding the properties of stars and obtaining age estimates that are as precise as possible, it may make sense to use as much information as is available.
However, to understand the assembly history of the Galaxy, it is ideal to have ages that are derived from information intrinsic to a star---Galactic information should be removed.

The primary contribution of this paper is to introduce a framework for making probabilistic predictions of the ages of stars.
Here we use the Galactic information in the form of \mgfe\ and \feh\ as well as stellar information in the form of \feh, \logg\, and \teff.
We then quantify how much age information comes from Galactic vs stellar models.
We introduce the methods used to do so---probabilistic random forests and SHapley Additive exPlanations---and demonstrate the framework's potential by applying it to a simple but realistic simulation of Milky Way stars.
We show that the model predictions are driven primarily by stellar evolution information, and as a result, we are able to make robust and interpretable predictions that reproduce expected trends in the data.

In Section \ref{sec:rc-sample} we repeat our analysis using a sample of APOGEE-mocked red clump stars.
This is a harder test case because the stars in the sample have very similar \logg\ values, as red clump stars lie in a small cluster on the Hertzsprung-Russell diagram.
Since \logg\ is be almost the same for all stars in that sample, we expect much less information to be present in the stellar models than for the Milky Way sample.
Thus, the models may need to obtain age information by proxying through chemical correlations, which are not intrinsic to the star itself.
We show that this is indeed the case, and discuss some implications of this result.
We also make an attempt to explicitly decorrelate the chemical information from the age information, and show how that process affects our results.
We note, however, that observers typically use [C/N] ratios rather than \logg\ and \teff\ to determine ages \citep[e.g.,][]{Salaris2015, Martig2016, Ho2017, Casali2019}.
In a follow-up paper, we will extend our framework to directly use stellar spectra rather than stellar parameters and abundances estimated from the spectra.

The layout of this paper is as follows.
In Section \ref{sec:data} we describe our data and the underlying Galactic and stellar models.
In Section \ref{sec:rf} we explain the probabilistic random forest regression model which we use to make probabilistic age predictions.
In Section \ref{sec:shap} we explain SHAP values, which allow us to quantify the relative importance of certain features in our predictive models, and then use them to determine the relative importance of features in our predictive model.
In Section \ref{sec:trends} we discuss how well the models are able to reproduce simulated age-abundance trend in the Galaxy, as a function of (1) choice of age estimator, (2) metallicity, and (3) whether there are chemical correlations.
In Section \ref{sec:rc-sample} we apply our framework to a harder test case of red clump stars where the model proxies age information using chemical correlations, and compare the results to those obtained from the Milky Way sample.
A summary of our findings is presented in Section \ref{sec:summary}.

\needspace{2\baselineskip}
\section{Mocking Milky Way Data}\label{sec:data}

\begin{figure}[htbp!]
  \centering
  \includegraphics[width=1\linewidth]{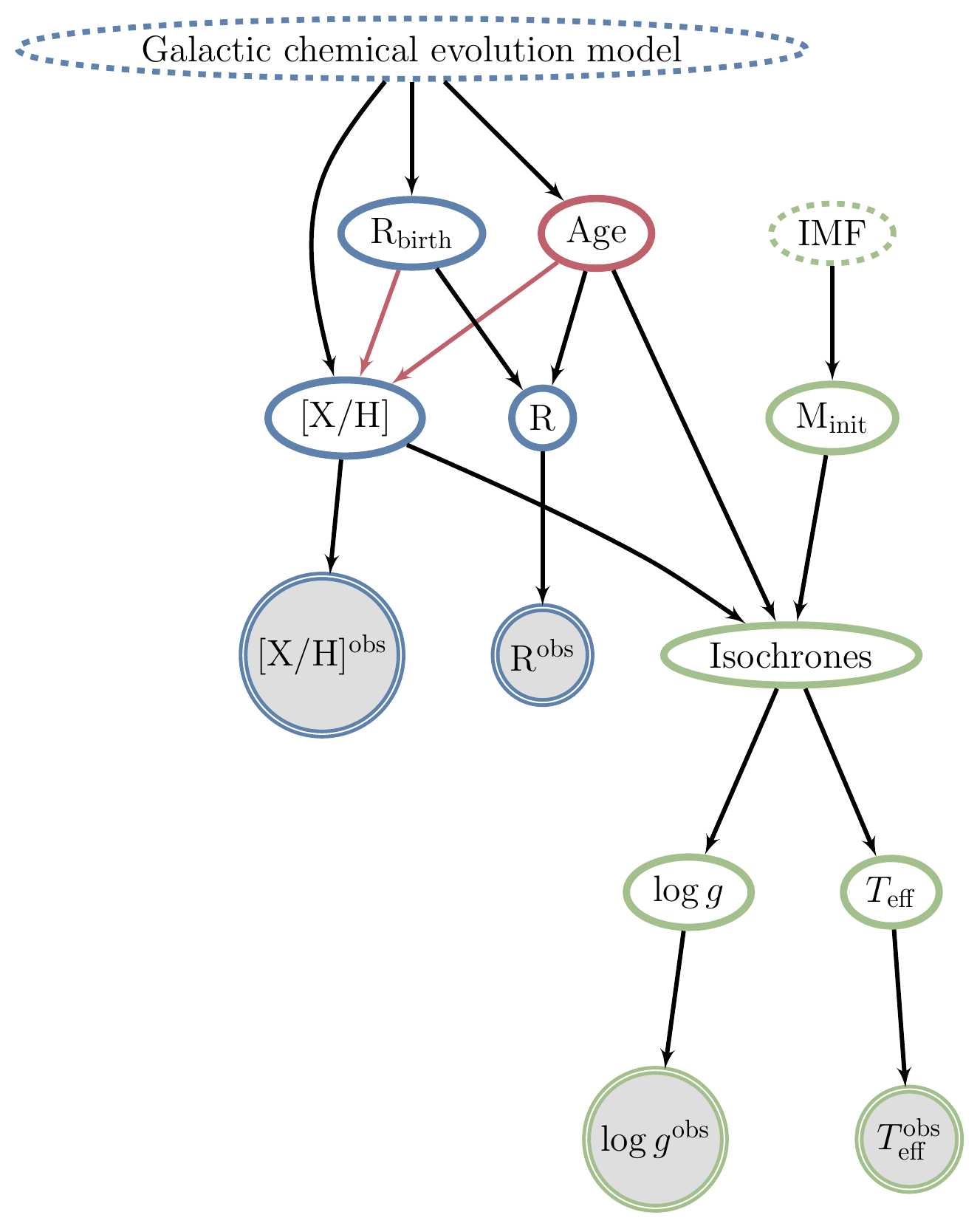}
  \caption{Graphical model showing the relation between variables in the dataset. Nodes with blue edges represent variables related to Galactic evolution, and nodes with green edges represent variables related to stellar evolution. Mock observations (i.e., with noise) are shown as nodes with two circles and grey shading. The nodes that are outlined with dotted lines represent models rather than random variables. Age, which is what we wish to predict, is shown in red. The red lines connecting [X/H] to R$_{\rm birth}$ and age are the correlations between chemical abundances and Galactic chemodynamical evolution, which we wish to remove.}
  \label{fig:dag}
\end{figure}

We produce a mock dataset of stars in the Milky Way, with known true ages, masses, and properties, and in Section \ref{sec:rf}, we train random forests to learn the masses from the properties.
The dataset contains stars of various evolutionary stages from the Milky Way's low-$\alpha$ disk.
The dataset contains age (mass), present-day Galactocentric radius, [Mg/H], [Fe/H], \logg\ and \teff.
The relationship between all the variables is represented in the graphical model in Figure \ref{fig:dag}.

We first sample from a variant of the \cite{Frankel2020} Milky Way's low-$\alpha$ disk model, to which we add [Mg/Fe] with a similar functional form as the [Fe/H]$(R_\mathrm{birth},t)$ described in their Eqs. 17 and 18, but with earlier enrichment\footnote{In particular, we use the same form with $\gamma_{\rm [Mg/H]} = 0.1$, $\nabla_{\rm [Mg/H]}= -0.065~{\rm dex\,kpc^{-1}}$, and ${\rm [Mg/H]_0} = 0.5~{\rm dex}$.}.
The model produces stars born from an inside-out star formation history, predicts with what abundances stars were born, and how they subsequently migrated from their birth places.
The [Mg/Fe] aspect of the model is not a best fit to the Milky Way but contains plausible correlations seen in real data, in particular faster enrichment in alpha elements than in iron-peak elements.

Given the Galactic variables, we then generate the stellar parameters using isochrones.
We begin by sampling initial masses plausible given the stars ages and metallicities from the \cite{Kroupa2001} initial mass function (IMF).
We consider evolutionary stages from the zero age main sequence (at equivalent evolutionary point, EEP, of 202 \citep{Choi2016}) to the thermally pulsating asymptotic giant branch (EEP = 808).
Beginning with a coarse EEP grid (EEP steps of 2.5), we use Brutus (Speagle et al., submitted to ApJ) to evaluate MIST isochrones \citep{Choi2016} over the EEP grid given some age and metallicity.
Here the isochrones are only a function of age and metallicity; Mg is not given.
This process yields stellar parameters for each value in the grid.
We select the set of parameters for which the initial mass most closely matches the initial mass sampled from the IMF.
We then perform a second pass using a finer grid (EEP steps of 0.05) around the EEP corresponding to the matched initial mass.
We again select the set of parameters for which the initial mass most closely matches the sampled initial mass, and use these parameters as the final stellar parameters.

Instead of working with stellar ages, we train our models to predict the stellar mass to avoid making assumptions about the mass loss prescription within our model itself. The age can be inferred from the mass using a number of methods afterward, because they are closely related by the evolutionary stage lifetime.
Here we use MIST isochrones \citep{Choi2016} to convert the predicted masses to ages.
However, we note that because isochrone interpolation is not exact, the conversion from mass to age introduces a small amount of error.
Thus, converting from age to mass and then back to age gives a result that differs from the original age by $\sim 2.5\%$.

The stellar parameters are correlated with both age and metallicity, since the isochrones were conditioned on these two variables.
Therefore, metallicity contains age information through both the Galactic model and the stellar model (isochrones).
In order to disentangle the effects of stellar ages from Galactic chemodynamical evolution, we also produce a ``control sample'' for which the ages are de-correlated from the chemical evolution model.
To do this, we randomly shuffle the ages produced by the Milky Way model, before using them to produce isochrones and train the random forests.
This shuffling step breaks the direct link between [X/H] and age (i.e., through the Galactic model).
In this control sample, the only age information coming from [X/H] should be indirect---purely from correlations with stellar evolution, through isochrones.

After generating all the parameters of interest, we simulate observational noise by applying Gaussian errors of 0.05 dex to [X/H], 0.1 dex to \logg, 2\% to \teff, and 0.04 dex to mass \citep{GarciaPerez2016, Majewski2017, Frankel2019}.

\begin{figure*}[htbp!]
  \centering
  \includegraphics[width=0.9\linewidth]{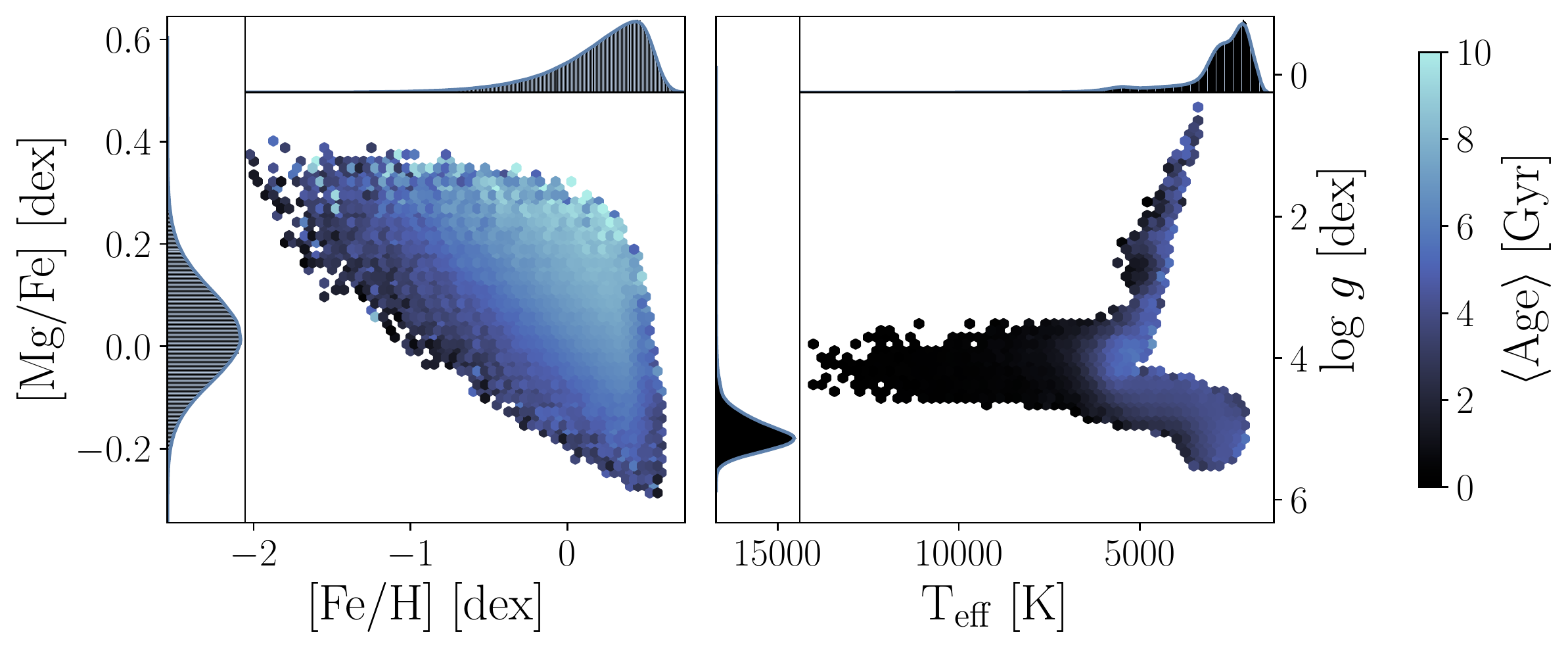}
  \caption{The simulated Milky Way sample. The left panel shows [Fe/H] against [Mg/Fe], with points colored by mean age within each hexagonal bin. The right panel shows the sample on a Hertzsprung-Russell diagram, where stellar parameters are generated conditional on ages and [Fe/H]. The points are again colored by mean age. All variables include observational noise. The inset axes in each of the panels show the marginal distributions.}
  \label{fig:data-mw}
\end{figure*}

In Figure \ref{fig:data-mw} we show our generated Milky Way sample.

\needspace{2\baselineskip}
\section{Probabilistic Masses (Ages) with Random Forests}\label{sec:rf}

With the mock dataset generated in Section \ref{sec:data}, we train a probabilistic random forest regression model (PRFR\footnote{A Python implementation of the PRFR model is available at \url{https://github.com/al-jshen/prfr}.}) to learn the masses from the stellar parameters and chemistry. The goal is to understand how PRFR is using the information from the training sample, and in particular to quantify how much chemical evolution it is using to learn stellar masses.

We split our dataset into a training set, a validation set, and a testing set, consisting of roughly 70\%, 20\%, and 10\% of the full dataset respectively. Model training is done with the training set, validation with the validation set, and model performance is determined using the testing set which is not seen during training.

The model that we use to learn masses from stars is a modified version of the popular random forest regression model \citep{Breiman2001}, which we call the probabilistic random forest regression model (PRFR).
The algorithm trains a number of decision trees, where each tree is given a bootstrapped and noisy version of the dataset.
Rather than aggregating (i.e., averaging) the predictions from all the trees as in a standard random forest regression model, we preserve all the individual predictions from the different trees.
This means that for each sample, we obtain n predictions, where n is the number of trees in the forest.
As a bootstrap can be thought of as a special case of the Bayesian bootstrap, which simulates a Bayesian posterior distribution \citep{Rubin1981, Hastie2009}, we can treat the samples from our bootstrapped trees as approximate samples from an posterior distribution for mass of the stars.

To account for measurement noise in our input parameters, we train each decision tree in the PRFR model using a draw from a Gaussian distribution centered around the measured values, and with standard deviation equal to the noise in the original dataset.
This allows us to, conditional on the noisy measured value, marginalize over the unknown true value.
Noise is also incorporated when making predictions in the same way.
During model training, we would also like to incorporate noise in the labels (i.e., the masses), which we do by using sample weighting.
Each sample is given an inverse variance weighting, so that stars with smaller errors in mass are weighted more heavily.

To ensure that the posterior (samples) are well-calibrated, we make predictions on the validation set and then compare the the masses in the validation set to the resulting mass PDFs.
Ideally, we want our PDFs to be calibrated so that the credible intervals match the coverage probability (i.e., roughly 50\% of the time, the true value lies in the 50\% credible interval).
We do this by finding a scaling factor to either narrow or broaden the PDF.
We then calculate the empirical CDF (ECDF) of the scaled PDF and evaluate it at the measured (``true'') mass.
Across multiple predictions, the resulting distribution of probability integral transform (PIT) values should be uniform in the ideal case \citep{Dawid1984,Diebold1998,Gneiting2007}.
We find the scaling value that minimizes the difference between the quantiles of the distribution of PIT values and a uniform distribution (i.e., makes the PIT distribution the closest to uniform).
We show in Appendix \ref{appendix:calibration} the impact of this procedure on the calibration of the posteriors.

Using the validation set, we also fit a multivariate linear regression model to the residuals, defined as the difference between the true values and the mean prediction, to explicitly correct for bias.
For predictions on the test set, we run the PRFR model (incorporating noise), adjust for bias, and apply the PDF calibration to obtain samples from a calibrated posterior for the masses.

\begin{figure*}[htbp!]
  \centering
  \includegraphics[width=0.85\linewidth]{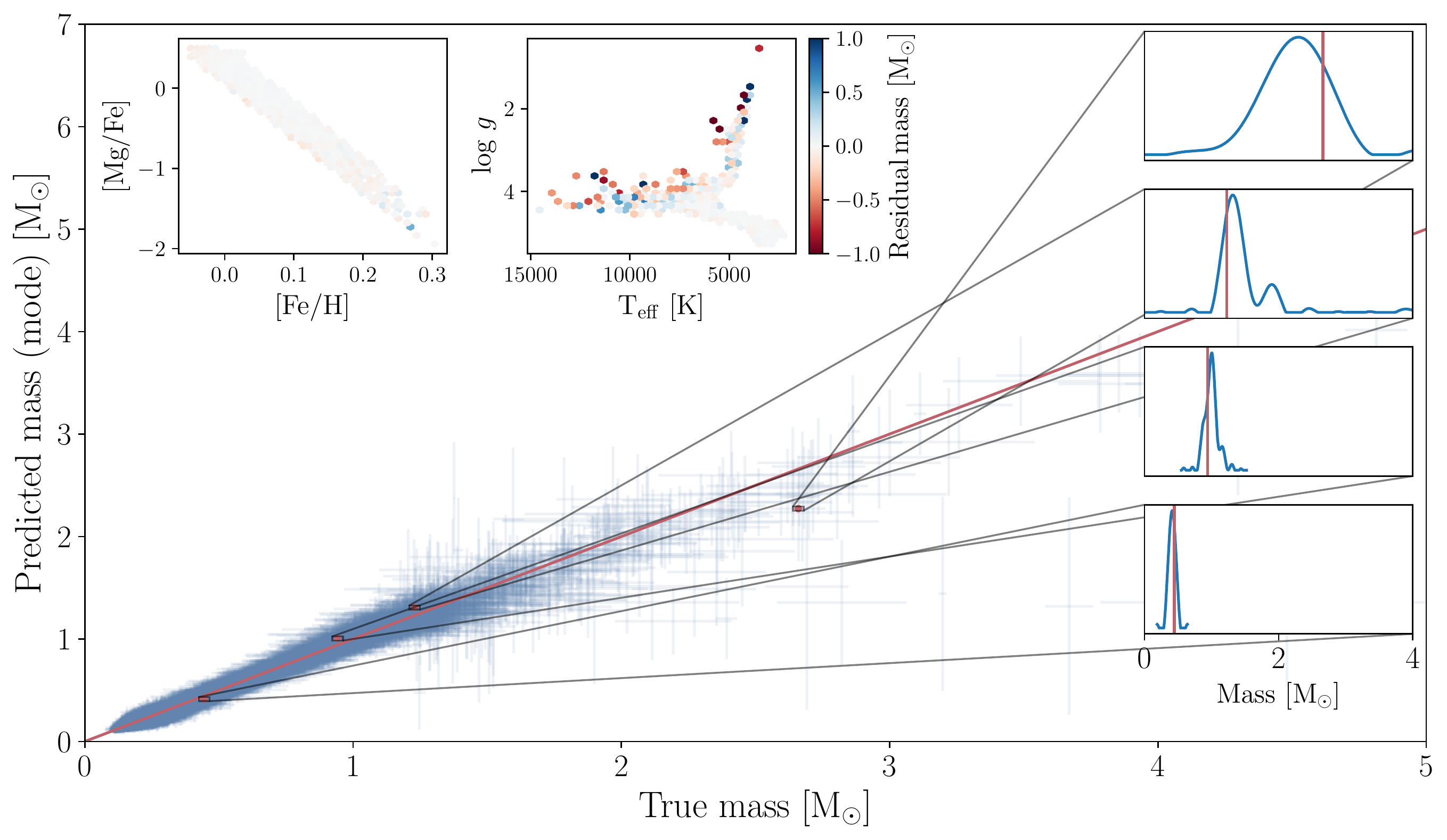}
  \caption{Predictions from PRFR model for Milky Way sample. The red diagonal line in the main panel is a 1:1 line. The blue crosses show the true mass and the mode estimate, with uncertainties corresponding to the simulated measurement error and the standard deviation of the posterior. In the top left, we show the residual mass as a function of different parameters. On the right, we show mass posteriors for several stars in blue, and the true mass for those stars as red vertical lines.}
  \label{fig:predictions_mw}
\end{figure*}

Figure \ref{fig:predictions_mw} shows the predicted mass against the true mass for a set of 25,000 mock stars from the sample described in Section \ref{sec:data}.
Here we use the modes of the predicted mass distributions for the predicted value, and the standard deviation of the distribution as the y-errorbar.
The difference between the noisy training mass and the true mass is used as the x-errorbar.

The PRFR model, across the entire mass range of the data, is able to accurately and precisely estimate the true mass, even when the training data are noisy.
The scatter in the residuals (defined to be the difference between the mode of the predicted mass distribution and the true mass) is $0.05~{\rm M_\odot}$, and the bias (by design of the model) is negligible at $0.006~{\rm M_\odot}$.

In the top left of the plot, we show the variation of residuals across parameter space.
At low logg, the stellar mass can be underestimated.
However, for most stars, the mode of the estimated posterior accurately traces the true mass of the stars.
On the right, we show examples of posteriors across a range of masses.
The blue lines represent kernel density estimates of the posteriors, and the red line shows the true mass.
In general, the posteriors are narrow and centered around the true mass.
That indicates that we are able to confidently recover the true mass, even when the training data are noisy.
The top right panel shows an exception where the mass is underestimated, but the posterior is appropriately wider, leading the mode and the mass to be separated by a reasonable distance.

\needspace{2\baselineskip}
\section{Interpreting Model Predictions with SHAP Values}\label{sec:shap}

When we use multiple variables in a model to make a prediction of age, how can we tell how much each of the variables contributes to our prediction?
SHapley Additive exPlanations (SHAP) values \cite{Lundberg2017} are a metric for quantifying the marginal contribution of different features in a model to that prediction.
They can be used to explain, for individual stars, how the inclusion of [Mg/Fe] (for example) in the model changes the prediction for the age of that star.
SHAP values are local, and the amount that a prediction changes due to conditioning on a particular feature depends on where in parameter space the star lies.

SHAP values have the desirable property of additivity; the total effect of multiple features is simply the sum of the effects of the individual features.
Consequently, the sum of the SHAP values of all the features gives the difference between the prediction from the full model (using all variables) and the baseline Naive Bayes prediction (i.e., randomly sampling from the age distribution), as well as what drives that difference.

\begin{figure*}[htbp!]
  \centering
  \includegraphics[width=1\linewidth]{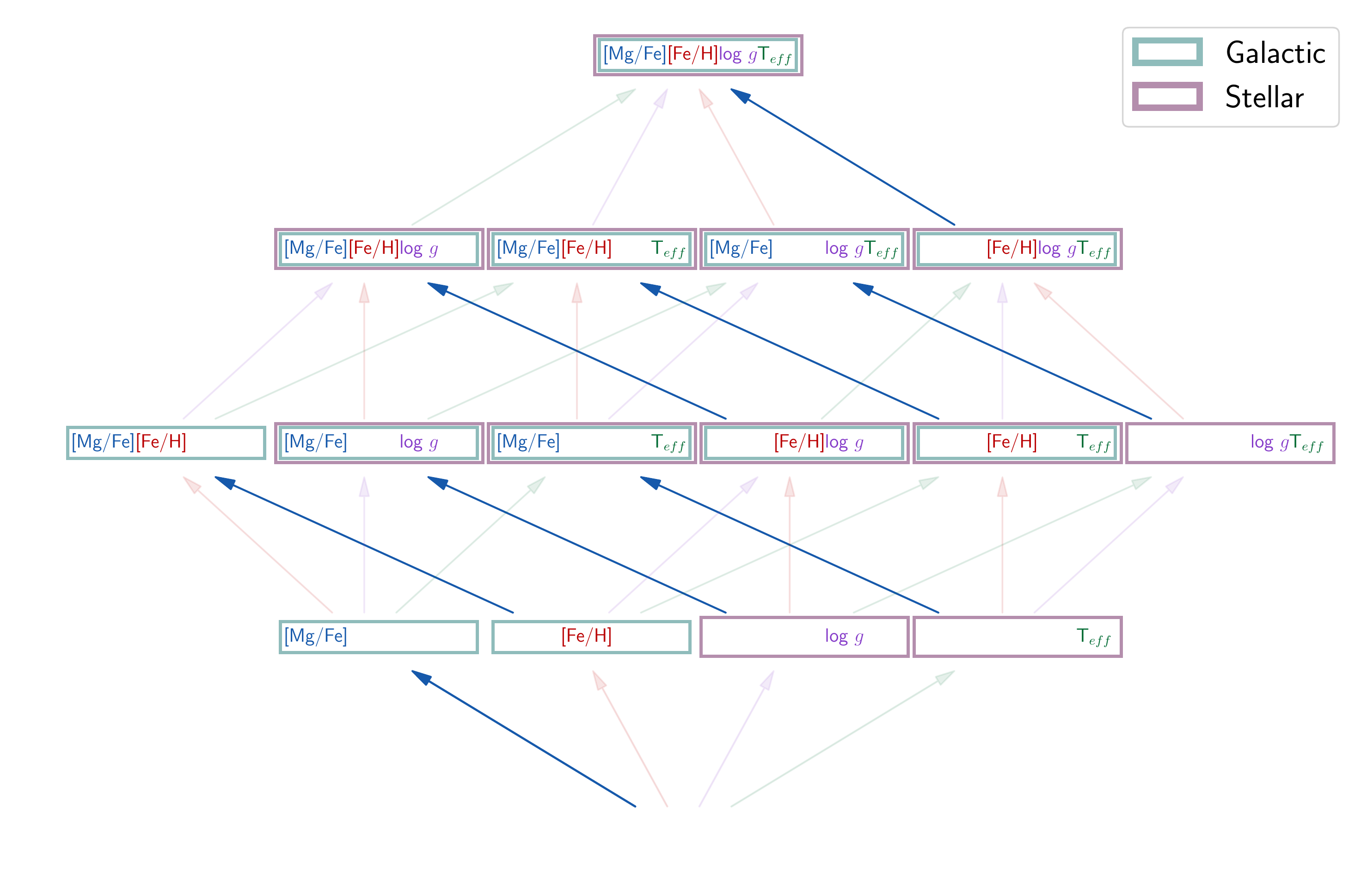}
  \caption{Graph of various models conditioned on different combinations of [Mg/Fe], [Fe/H], \logg\ and \teff. At the bottom of the plot is the baseline Naive Bayes model which is not conditioned on any features and makes predictions by randomly sampling from the age distribution. All of the other models are enclosed in colored rectangles, with teal indicating that the predictive model contains information from the Galactic evolution model and purple indicating that it contains information from the stellar evolution model. Arrows, representing model differences (i.e., going from a model without a particular feature to a model with that particular feature, all other features equal), are colored the same as the variables. For clarity we highlight the blue arrows, representing going from models without [Mg/Fe] to those with [Mg/Fe]. SHAP values for [Mg/Fe] are calculated as a weighted average of the differences represented by these arrows.}
  \label{fig:shapgraph}
\end{figure*}

Figure \ref{fig:shapgraph} shows a graph of models conditioned on various variables and indicates how SHAP values can be calculated.
We first calculate ``model differences'', which are the differences between models that include a particular feature and models that exclude that feature, all other features equal.
Each of these model differences is represented by an arrow, with the color of the arrow indicating the feature being considered for inclusion.
We focus on [Mg/Fe], indicated with blue arrows, for clarity.
Because the included feature might depend on other features, the SHAP value for a feature is a weighted average of all the relevant model differences.
The calculation of SHAP values requires that we have a separate model for each element in the power set of available features; for $K$ features, we need $2^K$ models---each feature is either included or not.

The formula for calculating SHAP values is
\begin{align}
  \text{SHAP}_j(x) = \sum_{\{s \in \mathcal{P}(F) \setminus \varnothing | j \in s\}} \frac{1}{|F| {|s| \choose |F|}} (f_s(x) - f_{\{s \setminus j\}}(x))
\end{align}
where $j$ is the feature being considered, $f$ is our model, $s$ is some set of features, $F$ is the set of all features, $x$ is the current point in parameter space where the SHAP value is being calculated, $\mathcal{P}$ denotes the power set, and $|s|$ is the cardinality of $s$.

In many cases, we would like to compare the relative importance of different features. To do so, we introduce normalized SHAP values, which are calculated as follows:
\begin{align}
  \rm SHAP_{N, j}(x) = \frac{SHAP_{j}(x)}{\sum_j |SHAP_{j}(x)|}
\end{align}
where $\rm{SHAP_N}$ is the normalized SHAP value and $j$ is the feature.
Simply, for a given object, the normalized SHAP value is the SHAP value of some feature divided by the sum of the absolute values of all the SHAP values for that object.
The interpretation of normalized SHAP values remains similar to that of SHAP values: if a normalized SHAP value is close to 0, that feature is not very informative, because adding or removing it does not change predictions much relative to the other features.
Conversely, if a value is close to -1 or 1, then that feature is almost entirely responsible for driving the prediction.
The magnitude of the normalized SHAP value is the relative importance of that feature, and the sign of the normalized SHAP value is the direction of the change in prediction.

\needspace{2\baselineskip}
\subsection{Application of SHAP Values to Probabilistic Age Predictions}\label{sec:shap-plots}

We train 16 PRFR models on our original mock dataset, and another 16 on the chemically decorrelated version; in each version, each of the 16 models is trained on a different subset of features.
SHAP values, as described in Section \ref{sec:shap}, are calculated using scalar predictions.
However, given that our models give us not only a single mass, but (samples from) a posterior distribution, we can instead calculate SHAP values using certain summary statistics (e.g., mean or median) of the posteriors.
We note that in theory, it is possible to calculate SHAP values for a grid of quantiles across the posteriors for each star, and that would allow us to get an idea of how the posterior distribution is influenced by the inclusion or exclusion of each of the variables.

Here we show results for the version of the Milky Way sample with observational noise.
For this simple test case of direct age estimation (see Section \ref{sec:where-ages} for justification), our model is robust even in the presence of noise, and so we do not show the test case without noise.

\begin{figure*}[htbp!]
  \centering
  \includegraphics[width=1\linewidth]{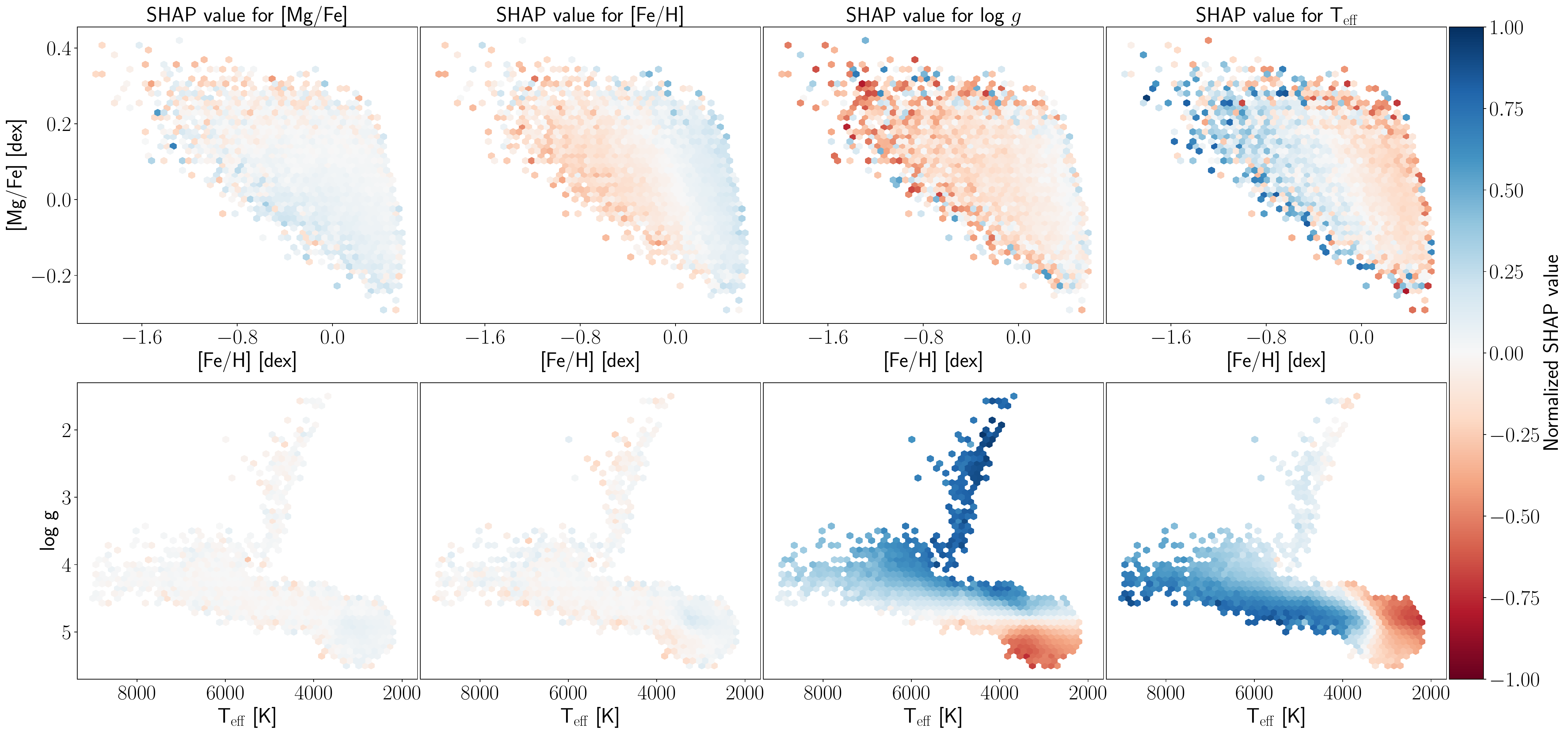}
  \caption{\textbf{Top row:} normalized SHAP values as a function of \feh\ and \mgfe. \textbf{Bottom row:} normalized SHAP values as a function of \teff\ and \logg. Each column represents the SHAP values for a different feature; moving from left to right, they are \feh, \mgfe, \logg, and \teff. The color scale indicates the magnitude of the normalized SHAP value. This plot is created with models trained on the noisy Milky Way data with chemical correlations.}
  \label{fig:shap2d_us_mw}
\end{figure*}

\begin{figure*}[htbp!]
  \centering
  \includegraphics[width=1\linewidth]{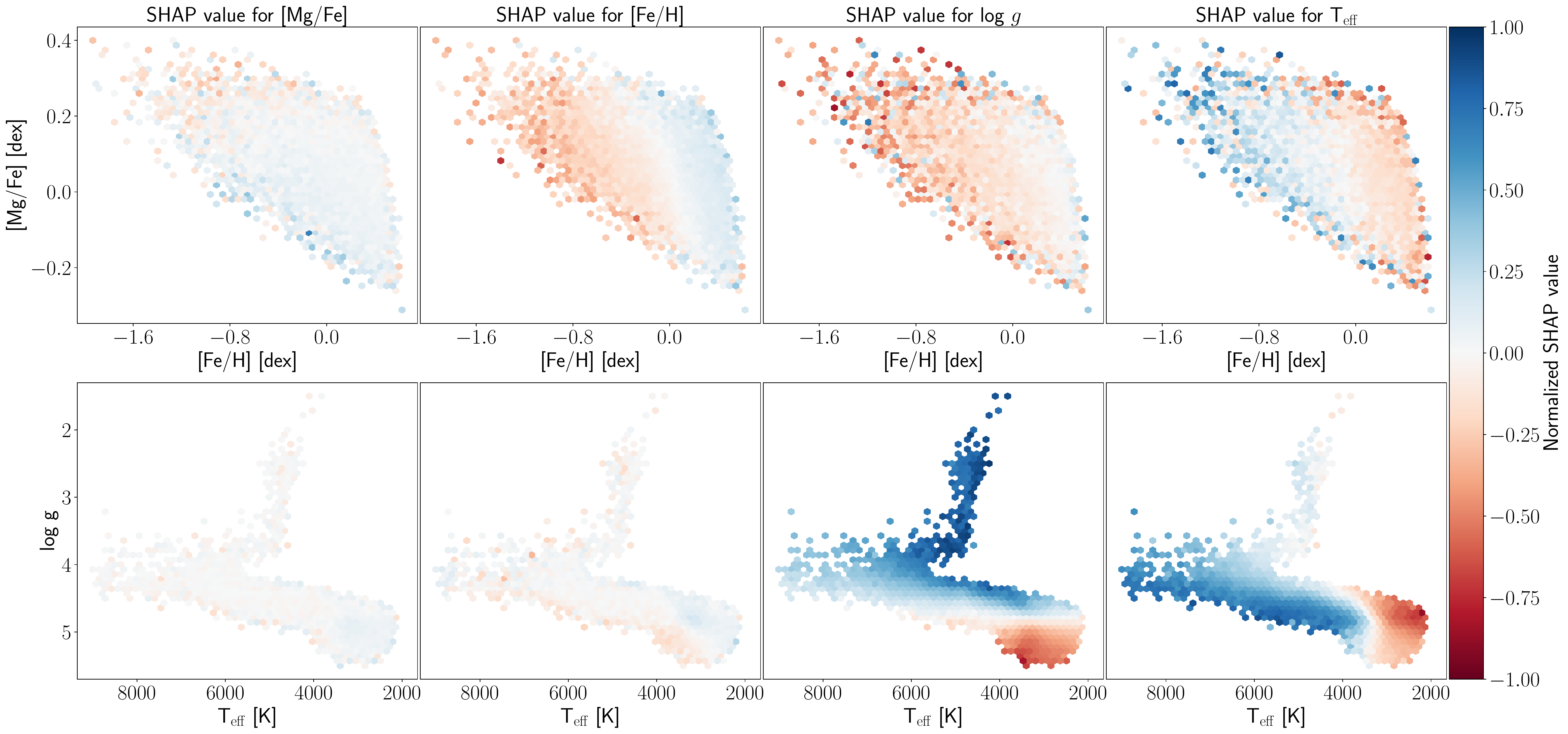}
  \caption{Same as Figure \ref{fig:shap2d_us_mw}, but with the noiseless Milky Way data without chemical correlations.}
  \label{fig:shap2d_s_mw}
\end{figure*}

In Figure \ref{fig:shap2d_us_mw}, we take the fiducial dataset (i.e., with chemical correlations) and predict the mass distributions of 1000 stars using all 16 models, and then for each star we calculate the SHAP values for each of the four parameters using the means of the mass distributions.
The figure shows how the normalized SHAP values vary across parameter space.
Each column represents the SHAP values for a different feature.
The top row shows the SHAP values as a function of \feh\ and \mgfe, and the bottom row shows the same values as a function of \teff\ and \logg.
This plot indicates the regions of parameter space where the inclusion of a feature increases or decreases mass predictions---relative to other variables---and by how much.
In Figure \ref{fig:shap2d_s_mw}, we show the same plot, but for the decorrelated dataset.

A priori, we expect the decorrelation process to reduce the importance of \mgfe\ and shunt the information into \feh.
This is because \mgfe\ becomes decorrelated from the ages and is thus mostly uninformative, except for the correlations that it retains through \feh.
\feh\ remains informative; in addition to the Galactic correlations that it has with age, it also has information about the ages through isochrones.

We see that for this test case, the difference in SHAP values between the fiducial and the chemically decorrelated datasets is minimal.
In particular, the SHAP values for \mgfe\ and \feh\ are (relatively) close to zero in all cases.
The results are similar even when Galactic correlations are explicitly removed; this is an indication that the models are directly learning ages and masses through stellar models, and not proxying them through Galactic correlations.
Indeed, the SHAP values for \logg\ and \teff\ are much larger than those for \feh\ and \mgfe---even in the presence of noise---indicating that they are more important for driving changes in model predictions.

\needspace{2\baselineskip}
\subsection{Where Is Age Information Coming From?}\label{sec:where-ages}

\begin{figure*}[htbp!]
  \centering
  \includegraphics[width=0.9\linewidth]{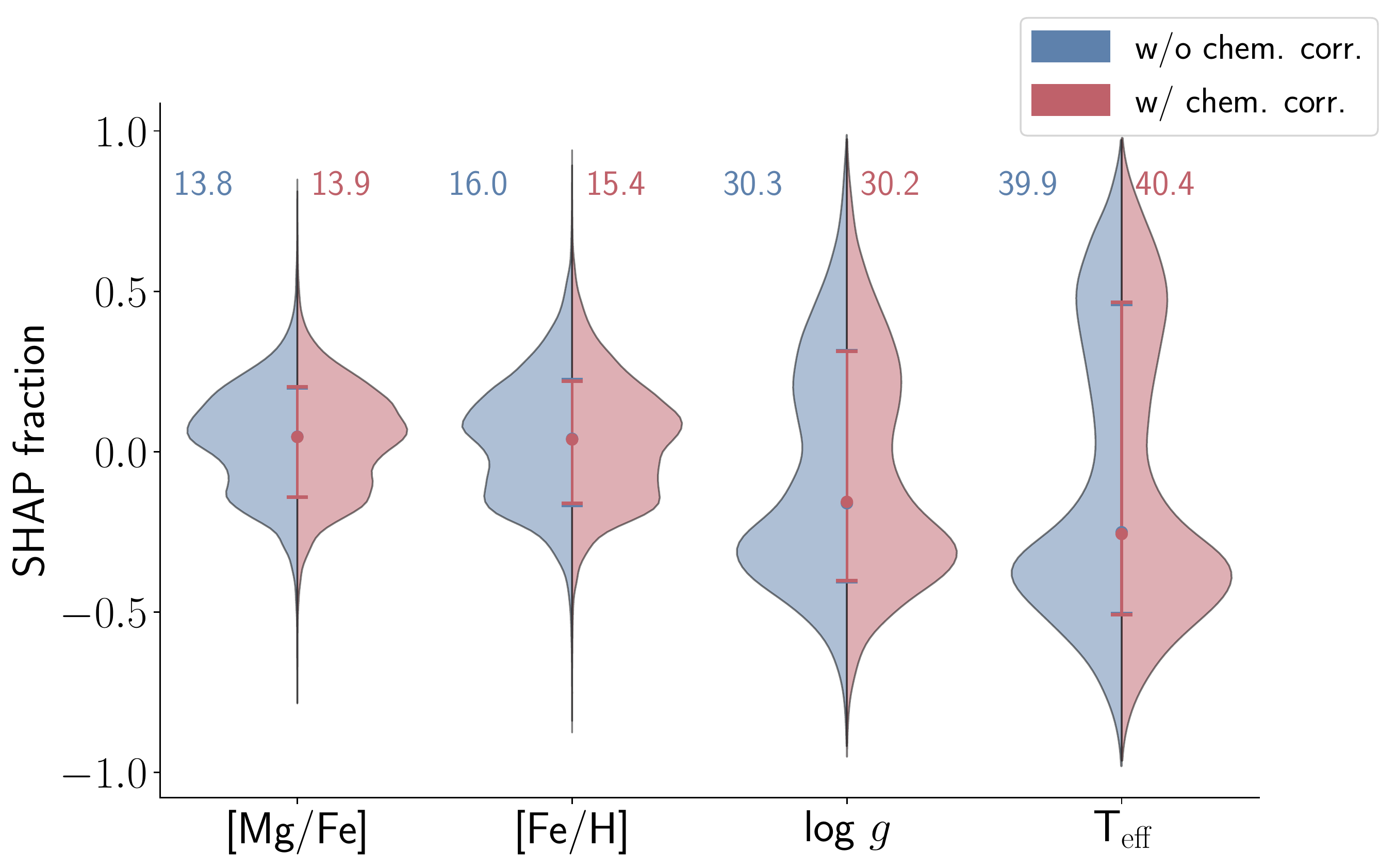}
  \caption{Split violin plot showing normalized SHAP values for the different parameters. The blue distributions on the left side of each violin represent the distribution of the SHAP values for the PRFR models trained on the fiducial dataset, and the red distributions represent the distribution of the SHAP values for the PRFR models trained on the data where chemical correlations are removed. The numbers above the violins are the means of the distribution of the absolute value of the normalized SHAP value. The dots in the middle of each violin represent the median of the distribution, and the errorbars show the 16th and 84th percentile. This plot is created with models trained on the noisy Milky Way dataset.}
  \label{fig:violin_mw_noisy}
\end{figure*}

As a way to more explicitly visualize the changes in SHAP values due to the decorrelation process, we create a split-violin plot in Figure \ref{fig:violin_mw_noisy} using the noisy Milky Way dataset.
Each of the violins shows the overall distribution of normalized SHAP values for the different model features;
the blue side shows the overall normalized SHAP values from the models trained on the fiducial dataset, and the red side shows the same but for the decorrelated dataset.
This gives us an overview of how the decorrelation process changes, overall, where information comes from.
The values above each of the violins are the means of the absolute value of the normalized SHAP values, which roughly gives the percentage contribution of information from that particular variable.

Confirming our previous results, we see that the SHAP values for \logg\ and \teff\ tend to be far from zero, indicating that they are relatively more informative than \mgfe\ and \feh\, for which the SHAP value distribution has the bulk of its mass around zero.

We note that the results shown here for the how well the PRFR models are able to predict ages using stellar models represents an ideal case where we have information about the mass loss prescription.
Thus, the models here likely have more information than we would in practice.
If we have a less complete understanding of how mass loss works, then the model might not be same as reality.
In that case, \logg\ and/or \teff\ might not be as informative, and there might be more information coming from chemical enrichment than is revealed by SHAP values.

\needspace{2\baselineskip}
\section{Age-Abundance Trends}\label{sec:trends}

\begin{figure*}[htbp!]
  \centering
  \gridline{%
    \fig{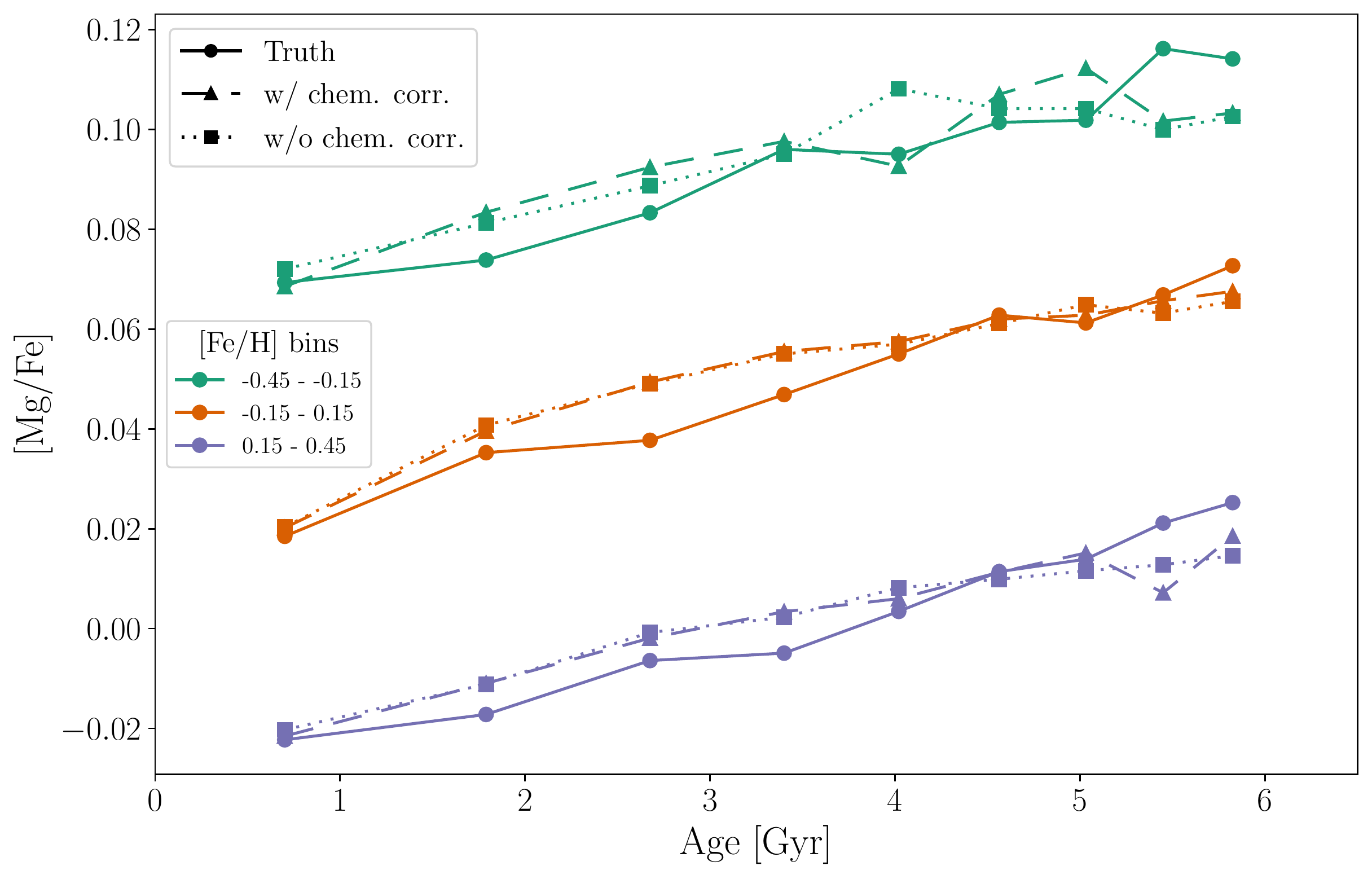}{0.48\textwidth}{(a)}
    \fig{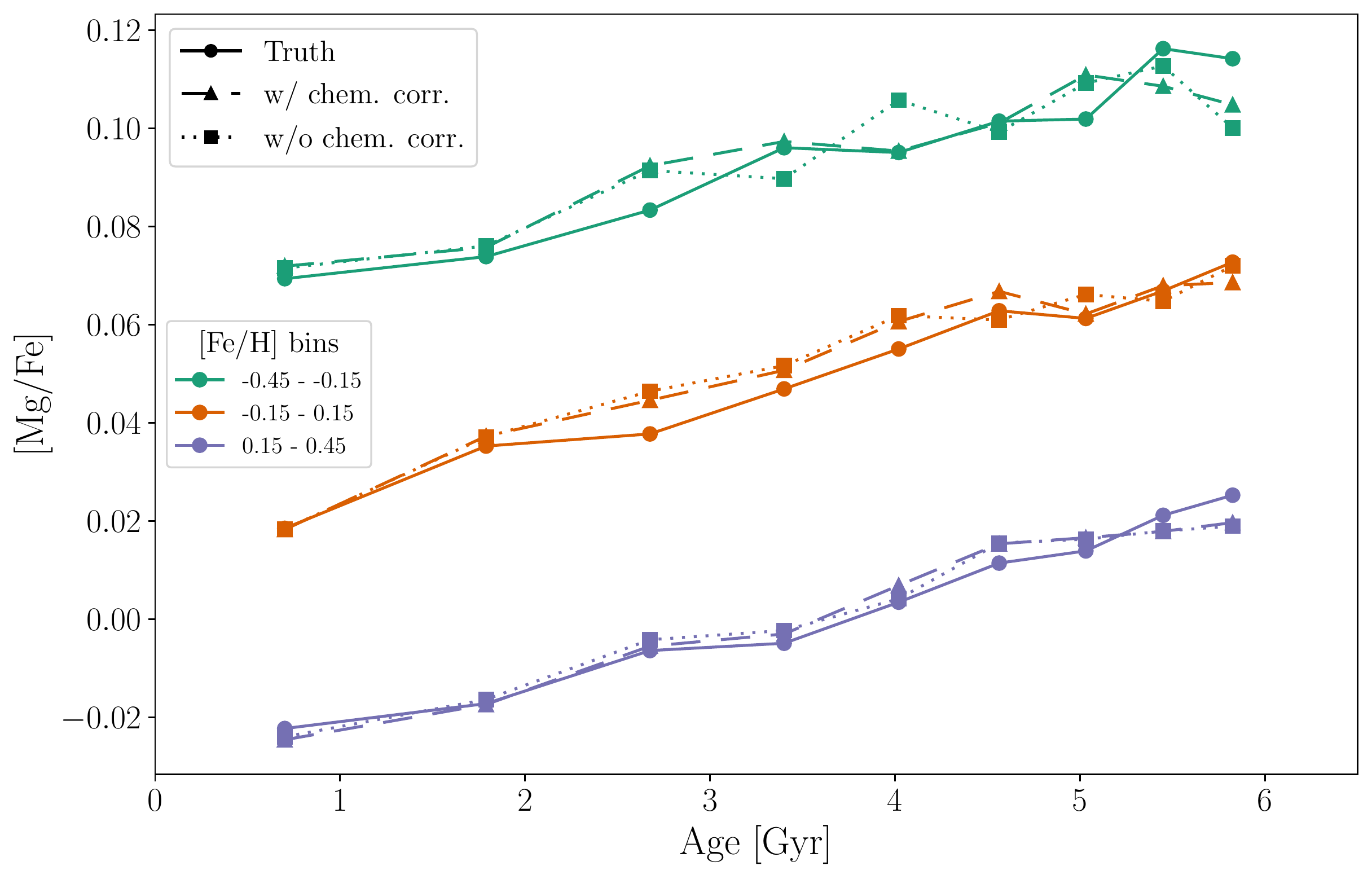}{0.48\textwidth}{(b)}
  }
  \caption{Trends in age and [Mg/Fe] across different metallicity bins, represented by the different colors. For each metallicity bin, we plot different age samples against [Mg/Fe] values: (1) the truth (i.e., the data) (2) the predictions from the PRFR model trained on the fiducial dataset, and (3) the predictions from the PRFR models trained on the chemically-decorrelated dataset. For each star, an age prediction is taken to be \textbf{(a)} a random sample from the conditional age distribution, or \textbf{(b)} the mode of the conditional age distribution. This gives us a sample of (age, [Mg/Fe]) pairs, which we then bin in age. We then use the center of each age bin and the median of all [Mg/Fe] values in that age and metallicity bin to create a line.
  }
  \label{fig:trends_mw}
\end{figure*}

We would like to examine whether the models are able to accurately reproduce age-abundance trends in the data.
In Figure \ref{fig:trends_mw}, we show the age-abundance trends for the Milky Way sample.
On the x-axes, we show the age of the star, and on the y-axes, we show the \mgfe\ value from the dataset.
In each panel, we consider stars in three different \feh\ bins, as indicated by the color of the lines.
Within each bin, there are three sets of lines of identical colors: one created with the true age, one created with the predictions from the model trained on the fiducial dataset, one created with the predictions from the model trained on the dataset with chemical correlations removed.
That is, each line is created from datapoints with the same \mgfe\ values, but differ in their ages (x-positions).

Recalling that the outputs of our PRFR model are samples from a posterior, we can consider how taking different point estimates from the posterior can influence the predicted age-abundance trends.
On the left, we consider random samples from the conditional age distributions (converted from the conditional mass distributions).
On the right, we consider the mode from the conditional age distributions.
We see that the observed trends are very similar between the different metallicity bins, and also between the two panels.
In addition, the lines corresponding to the model predicted ages and the true ages agree well.
This shows that in this test case where we are able to directly estimate the ages without proxying through chemical correlations, the predictions from the model---regardless of whether we take random samples (i.e., sample the entire posterior) or a summary statistic that traces the bulk of the distribution---we are able to accurately reproduce age-abundance trends.

\needspace{2\baselineskip}
\section{Red Clump Sample}\label{sec:rc-sample}

We repeat our entire analysis using an APOGEE-mocked red clump sample of stars. We generate this dataset following the method described in Section \ref{sec:data}, but using EEPs ranging from 631 to 707, corresponding to the core helium burning phase of a star's lifetime.

\begin{figure*}[htbp!]
  \centering
  \includegraphics[width=0.9\linewidth]{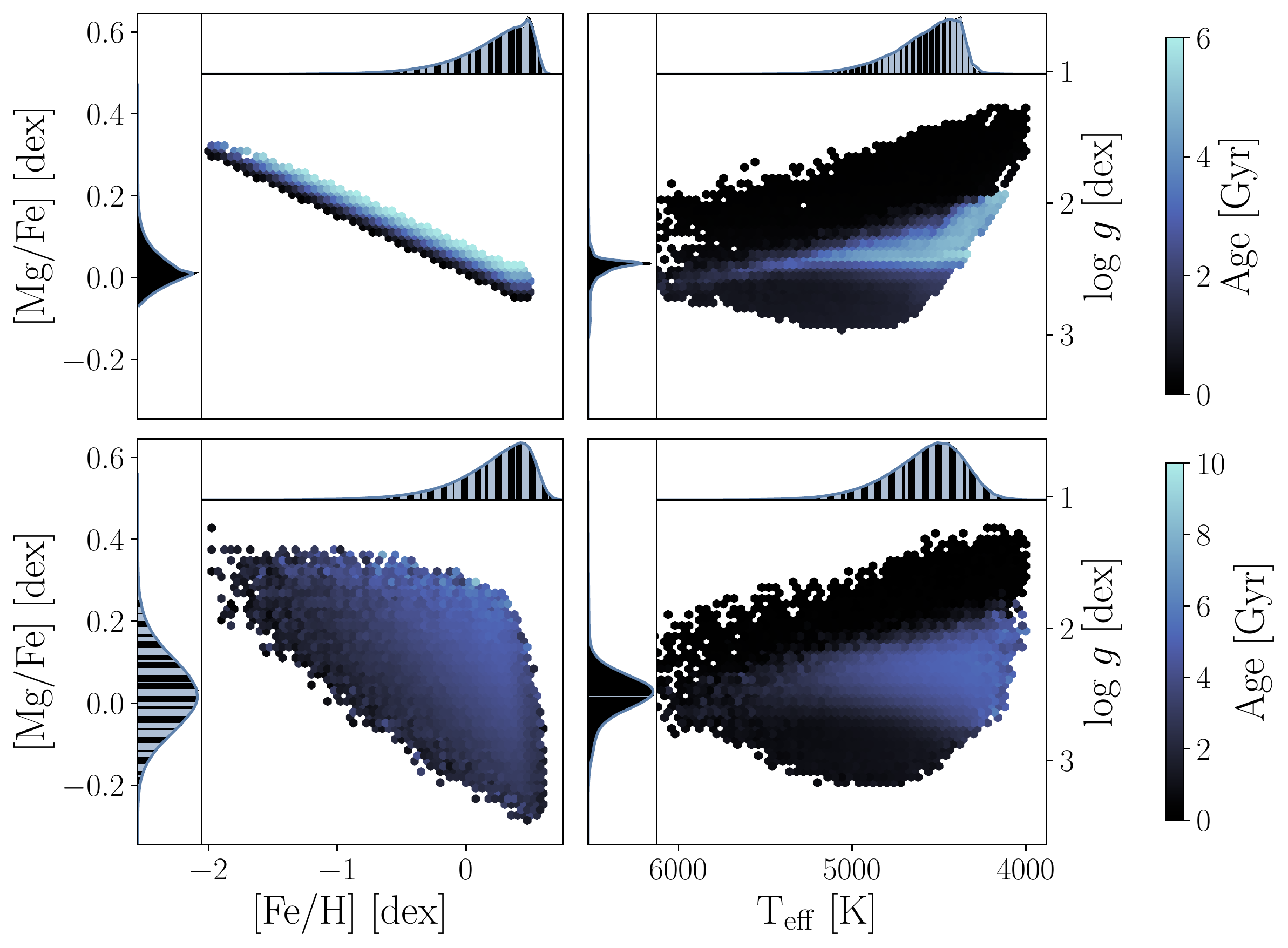}
  \caption{The simulated APOGEE-mocked red clump sample. The two plots in the left column show [Fe/H] against [Mg/Fe], with points colored by age. The plots in the right column show the sample on a Hertzsprung-Russell diagram, where stellar parameters are generated conditional on ages and [Fe/H]. The points are again colored by age. The top two plots does not include noise, and shows the true underlying trend from the simulation; the bottom two plots and the (corresponding colorbar) show the noisy parameter values and ages. The inset axes in each panel show the marginal distributions.}
  \label{fig:data-apogee}
\end{figure*}

Figure \ref{fig:data-apogee} shows the red clump sample in parameter space.
The overall distributions of \mgfe\ and \feh\ in the red clump sample is identical to those in the Milky Way sample, but the \logg\ and \teff\ values span a much smaller range.
In particular, the \logg\ range for the red clump sample, especially in the absence of observational noise, is extremely small, with most stars having $\log g \approx 2.5$.

This particular sample is a test case where the information content coming from stellar evolution should be much lower, precisely due to the highly constrained range in values as explained above---in the limiting case, we would be using a constant as one of the variables, which should be totally uninformative.
As a consequence, the model tries to learn to predict masses using the available chemical proxies, rather than only through stellar evolution.
We stress, however, that it is indeed possible to use stellar evolution to predict masses for red clump stars; there is simply less stellar evolution information available.

\needspace{2\baselineskip}
\subsection{Mass (Age) Estimates with PRFR}\label{sec:analysis-rc}

\begin{figure*}[htbp!]
  \centering
  \includegraphics[width=0.85\linewidth]{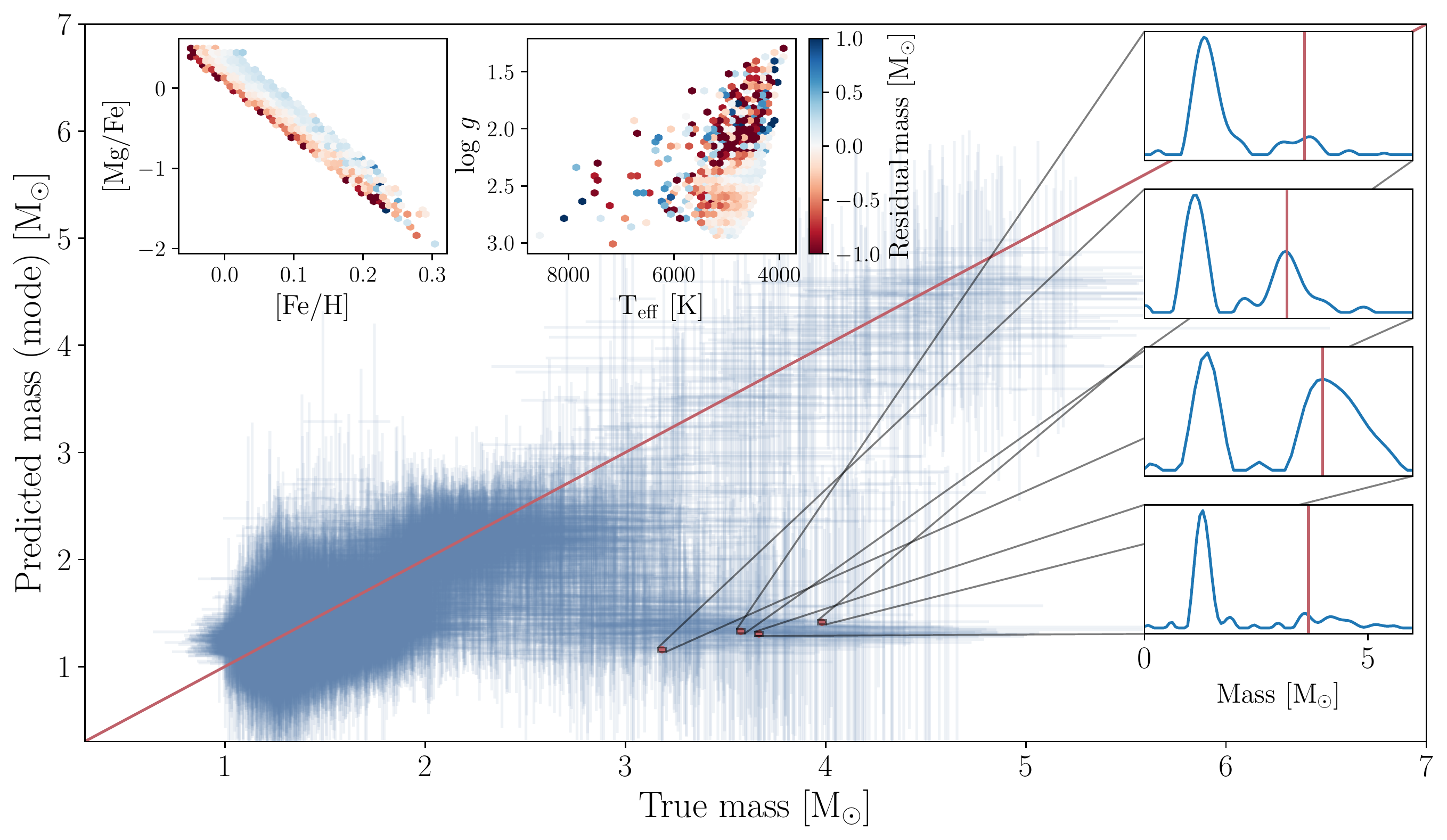}
  \caption{Predictions from PRFR model for RC sample. The inset panels on the top left show how the residuals vary across parameter space; we find that there is structure in the residuals. The inset panels on the right show particular cases of poor predictions. In each inset panel the blue line shows a kernel density estimate of the predicted mass distribution, and the red vertical line shows the true mass. In all four cases the true mass is a secondary solution. }
  \label{fig:predictions_apogee}
\end{figure*}

Figure \ref{fig:predictions_apogee} shows a simple one-to-one comparison of the true mass with the mass predictions from the PRFR model for the RC sample.
In general, the model prediction is in agreement with the true mass (but less so than for the full sample); the scatter is $0.33~{\rm M_\odot}$.
However, there is a particular horizontal clump of stars where the PRFR model systematically underpredicts the mass to be $\sim 1.2~{\rm M_\odot}$, even in cases where the true mass is $\sim 4~{\rm M_\odot}$.

Torward the right side of the plot, we randomly select four stars from this clump and show their predicted mass distributions as inset axes.
We see that in all cases, there is a clear peak in the posterior near $\sim 1.2~{\rm M_\odot}$ which is taken to be the prediction in the main scatterplot, as well as a secondary solution at higher masses which matches the true mass.
This is an indication that point predictions are insufficient when trying to reproduce the full complexity of the mass distribution in the data; many of the posteriors are not, e.g., simple Gaussians, and thus the full posterior distribution should be taken into account.

We further examine the performance of the PRFR model in the top-left inset axes of Figure \ref{fig:predictions_apogee}.
The two panels show the residuals of the model across parameter space.
We find that there is a diagonal residual gradient across \mgfe\ and \feh\ space, with masses at lower \mgfe\ and \feh\ being underpredicted.
On the right panel, there is a red-colored clump of stars around ${\rm T_{eff} \approx 4700~{\rm K}}$ and ${\log g \approx 2}$ where the residuals are mostly negative, indicating that the PRFR model systematically underpredicts the mass.
These stars are likely the same as the ones in the clump in the main part of the figure.

\needspace{2\baselineskip}
\subsection{SHAP Values}\label{sec:shap-apogee}

\begin{figure*}[htbp!]
   \centering
   \includegraphics[width=1\linewidth]{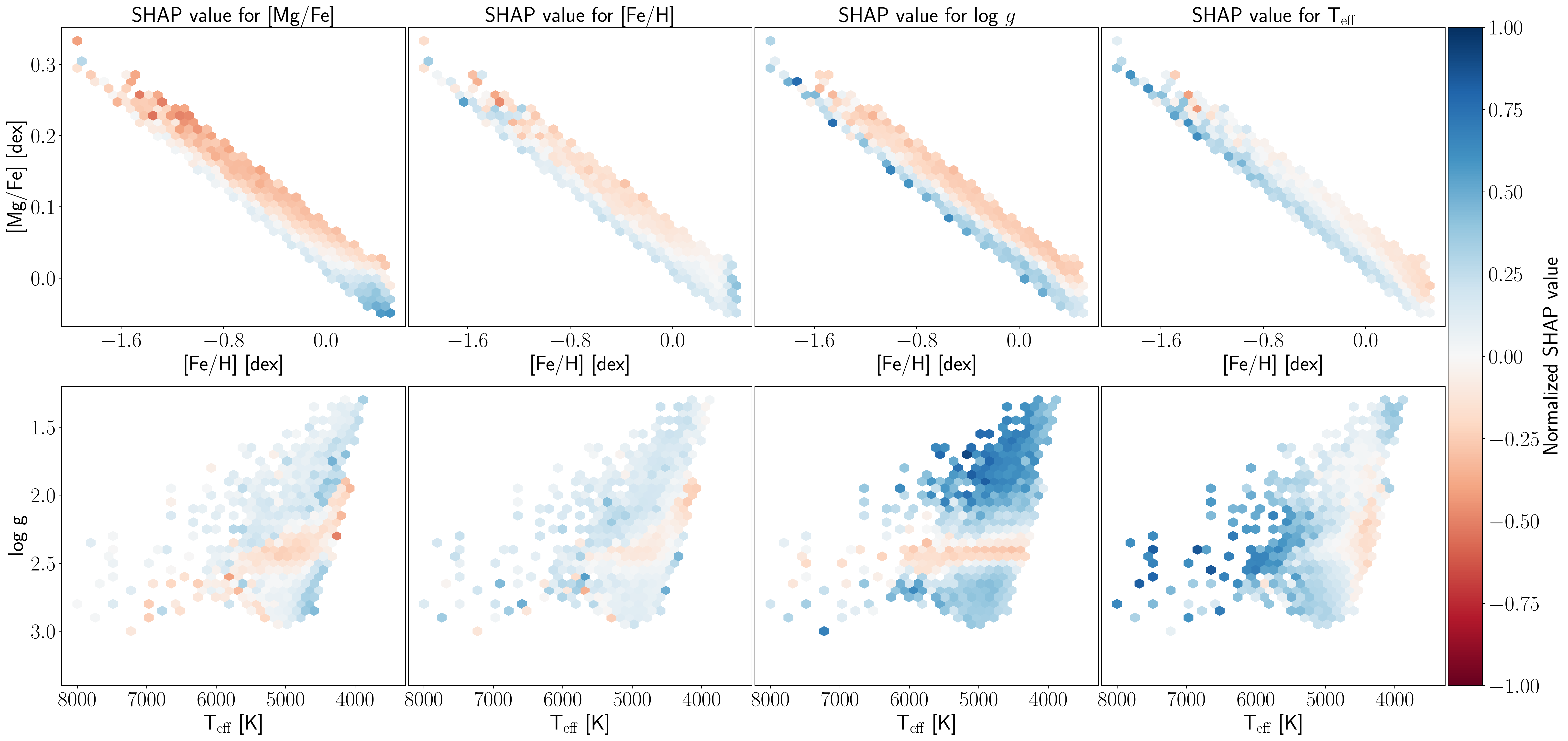}
   \caption{Same as Figure \ref{fig:shap2d_us_mw}, but with the noiseless red clump data with chemical correlations.}
   \label{fig:shap2d_us_apogee_noiseless}
 \end{figure*}

\begin{figure*}[htbp!]
   \centering
   \includegraphics[width=1\linewidth]{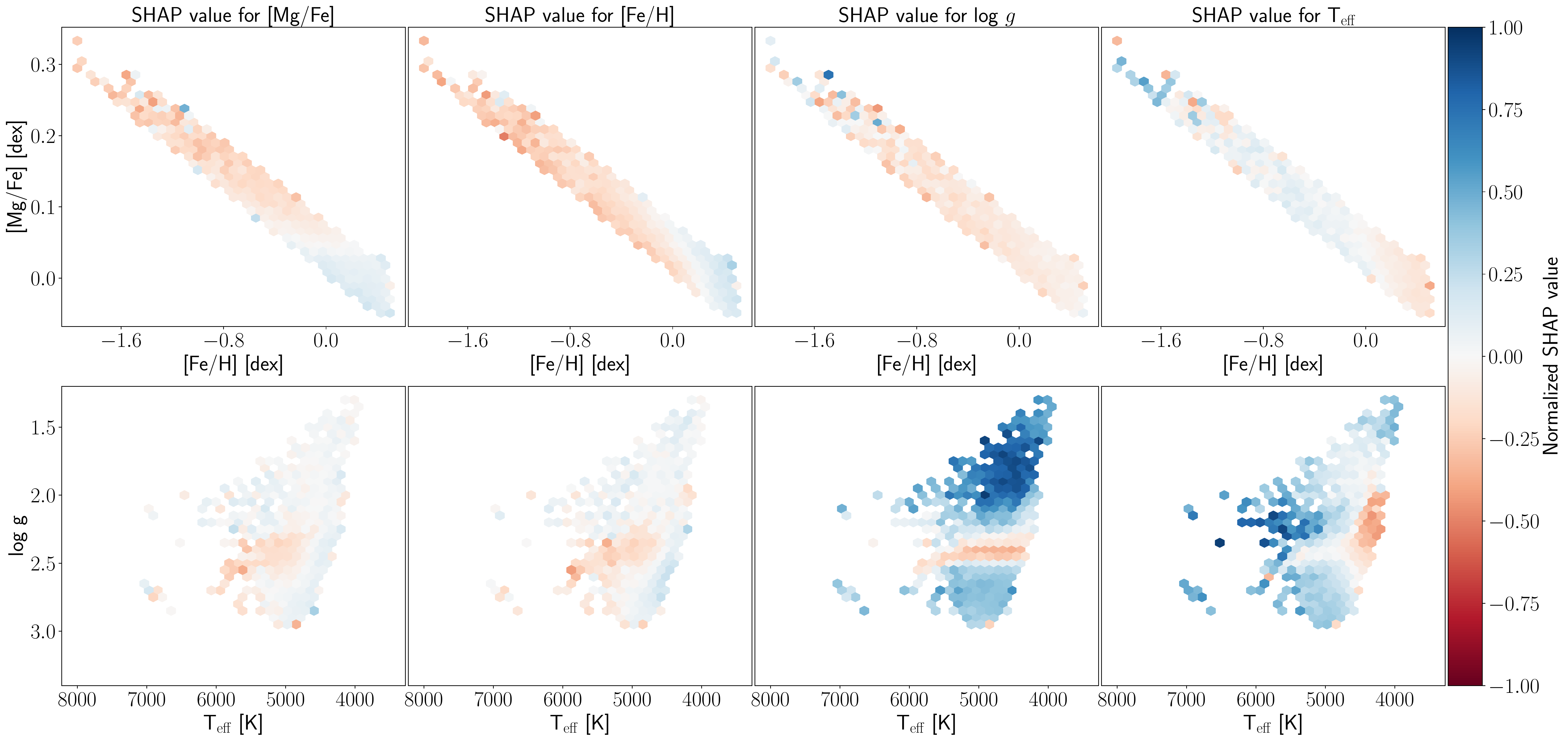}
   \caption{Same as Figure \ref{fig:shap2d_us_mw}, but with the noiseless red clump data without chemical correlations.}
   \label{fig:shap2d_s_apogee_noiseless}
 \end{figure*}

\begin{figure*}[htbp!]
  \centering
  \includegraphics[width=1\linewidth]{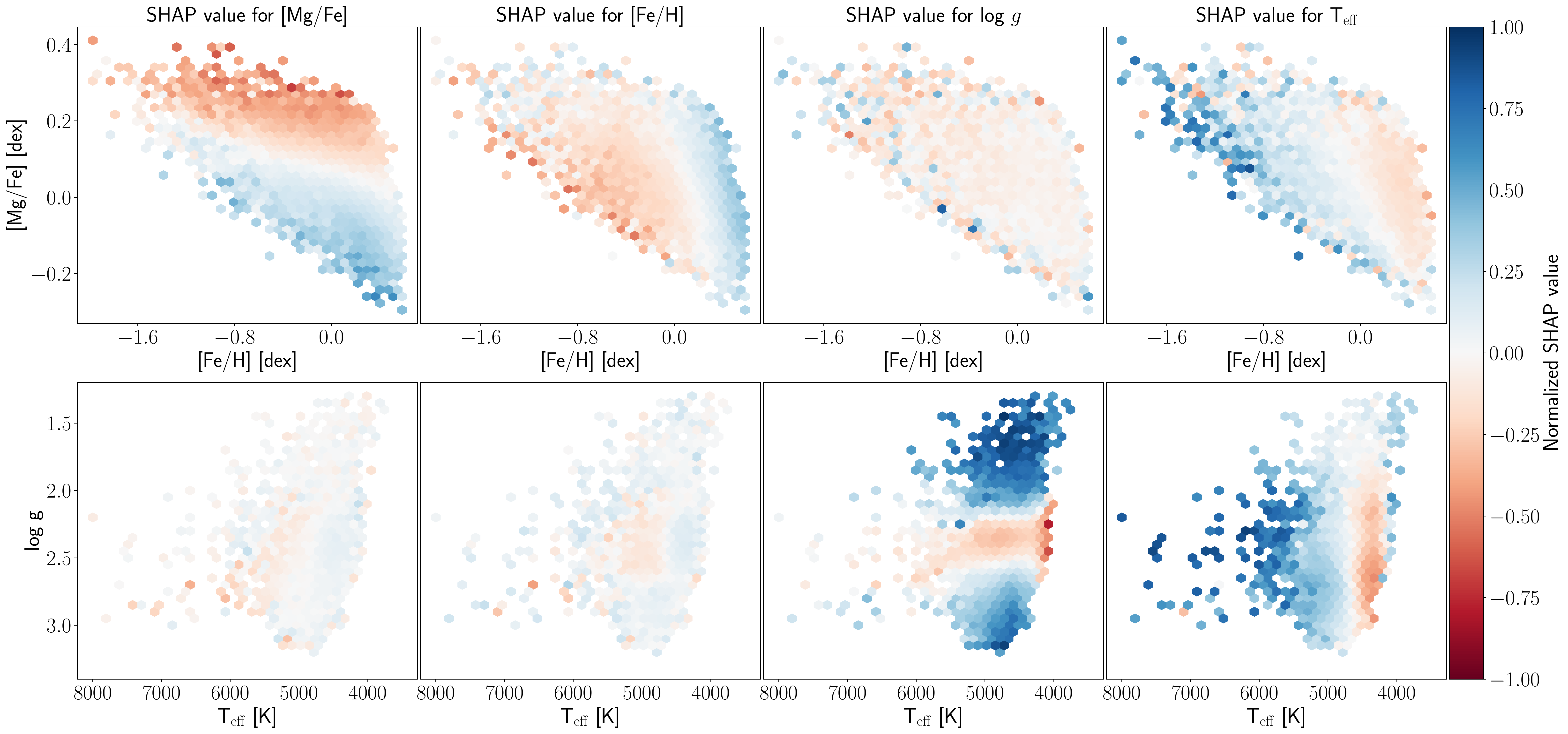}
  \caption{Same as Figure \ref{fig:shap2d_us_mw}, but with the noisy red clump data with chemical correlations.}
  \label{fig:shap2d_us_apogee}
\end{figure*}

\begin{figure*}[htbp!]
  \centering
  \includegraphics[width=1\linewidth]{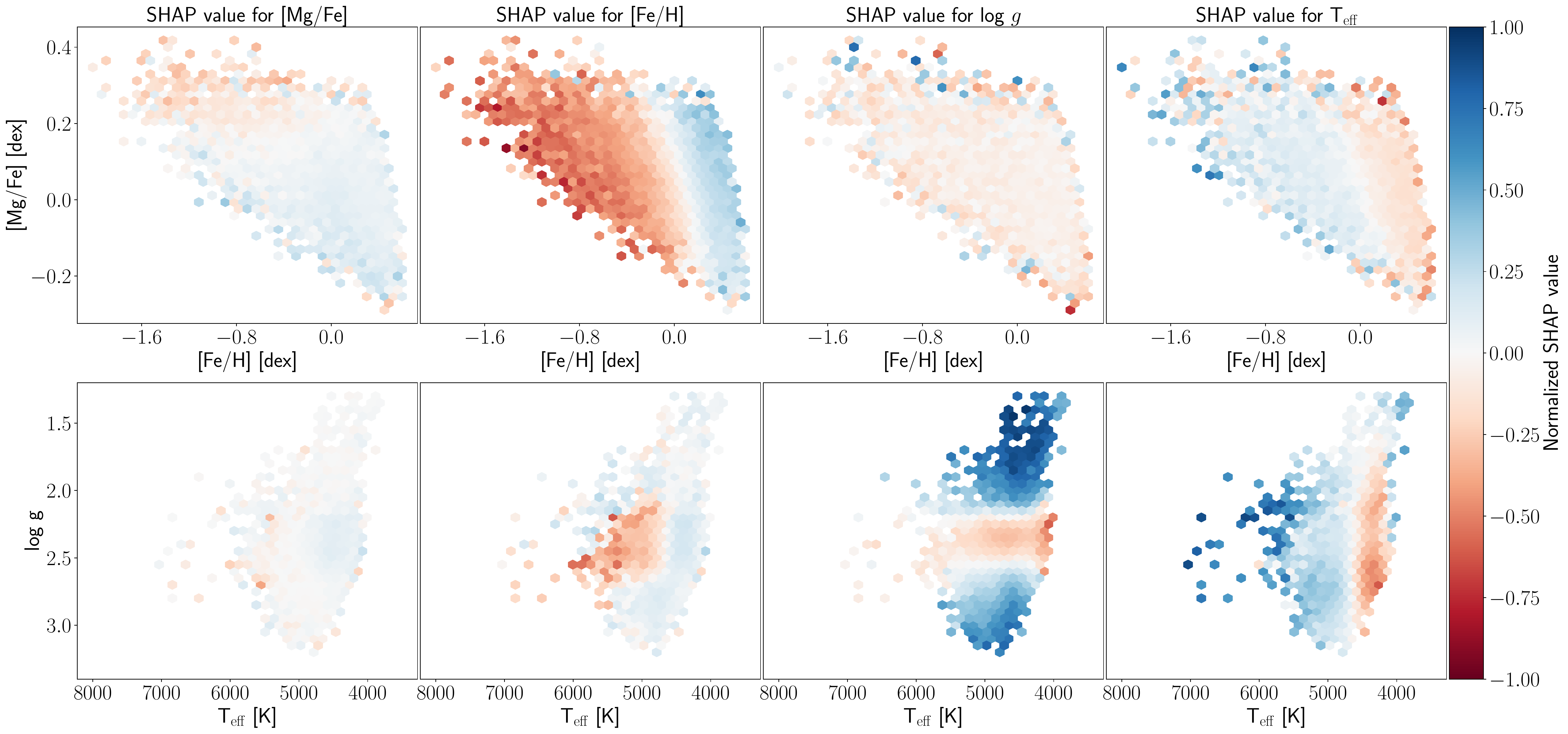}
  \caption{Same as Figure \ref{fig:shap2d_us_mw}, but with the noisy red clump data without chemical correlations.}
  \label{fig:shap2d_s_apogee}
\end{figure*}

Here we recreate the 2D SHAP value plots from Section \ref{sec:shap-plots} using the RC sample.
Figures \ref{fig:shap2d_us_apogee} and \ref{fig:shap2d_s_apogee} are created from the noisy RC datasets with and without chemical correlations, respectively.
By comparing the results from the fiducial dataset to those from the chemically decorrelated dataset, we can identify the effect that the decorrelation process has on the model predictions.
 In the top row of Figure \ref{fig:shap2d_us_apogee_noiseless}, we see that there are diagonal gradients in all panels; this indicates that the SHAP values for all four variables depends on chemistry. 
 When chemical correlations are removed in Figure \ref{fig:shap2d_s_apogee_noiseless}, the diagonal gradient effectively disappears in of the top panels, and the SHAP values are essentially independent of the combined chemistry, as expected.
 There is still a weak horizontal gradient in the upper right panel, indicating that the \teff\ SHAP value varies with metallicity, but this is reasonable since we expect the two to be covariant \citep{Rix2022}.

Compared to the Milky Way sample, the \mgfe\ and \feh\ SHAP values for the RC sample are far less clustered around zero.
There are clear gradients in the left two panels in the upper row of Figure \ref{fig:shap2d_us_apogee}; the magnitudes of the SHAP values there are comparable to those in the right two panels in the upper row, indicating that the normalized SHAP values are much more balanced across the four variables.
In constrast, in Figures \ref{fig:shap2d_us_mw} and \ref{fig:shap2d_s_mw}, the \mgfe\ and \feh\ SHAP values are clustered around zero, as indicated by the near-white colors in the four left-most panels in each of the plots.

The results from the RC sample show that the SHAP values for the chemically decorrelated dataset are somewhat similar to those for the fiducial dataset, particularly for \logg\ and \teff.
However, the SHAP values for \feh\ and \mgfe\ are clearly different; the strong gradient in the SHAP value for \mgfe\ (seen in the upper left panel of Figure \ref{fig:shap2d_us_apogee}) essentially disappears in Figure \ref{fig:shap2d_s_apogee}.
At the same time, the magnitude of the SHAP values for \feh\ are larger in the decorrelated version; especially for low \feh, the SHAP values in Figure \ref{fig:shap2d_s_apogee} are closer to -1 than those in Figure \ref{fig:shap2d_s_apogee}.
These two results, in combination, suggest that information content from \mgfe\ has been shunted into \feh\ in the model.

\needspace{2\baselineskip}
\subsection{Where Is Age Information Coming From?}\label{sec:where-age-apogee}

The SHAP value plots suggest that \feh\ and \mgfe\ provide more information for the red clump sample than for the Milky Way sample.
While stellar evolution remains quite informative, chemical correlations play a much larger role in driving predictions.
The decorrelation process shifts the bulk of the distribution of SHAP values for \mgfe\ to be close to zero, indicating that in most cases, \mgfe\ becomes less responsible for driving predictions than when chemical correlations are present.
At the same time, \logg\ and \teff\ become slightly more informative, as the bulks of the distributions shift slightly further away from zero in the negative direction.

When observational noise is included, we see that the decorrelation process has a similar effect on \mgfe\ as before.
However, \feh\ now becomes much more informative.
When we remove chemical correlations, we eliminate any direct age information contained in \mgfe; the only age information contained in the decorrelated \mgfe\ should be the correlation that it has with \feh, which still contains age information because of isochrones.

\needspace{2\baselineskip}
\subsection{Age-Abundance Trends}\label{sec:trends-apogee}

\begin{figure*}[htbp!]
  \centering
  \gridline{%
    \fig{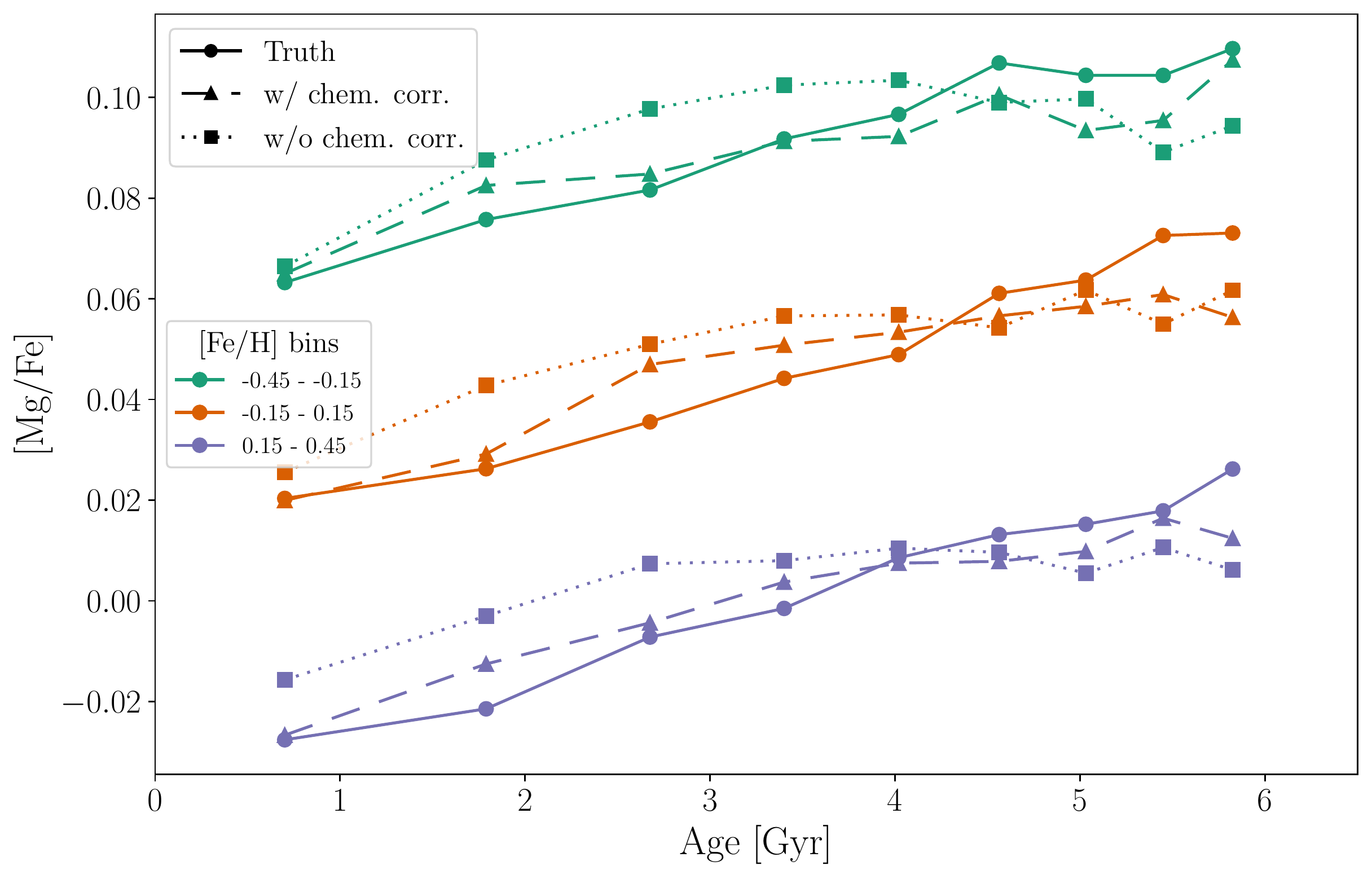}{0.48\textwidth}{(a)}
    \fig{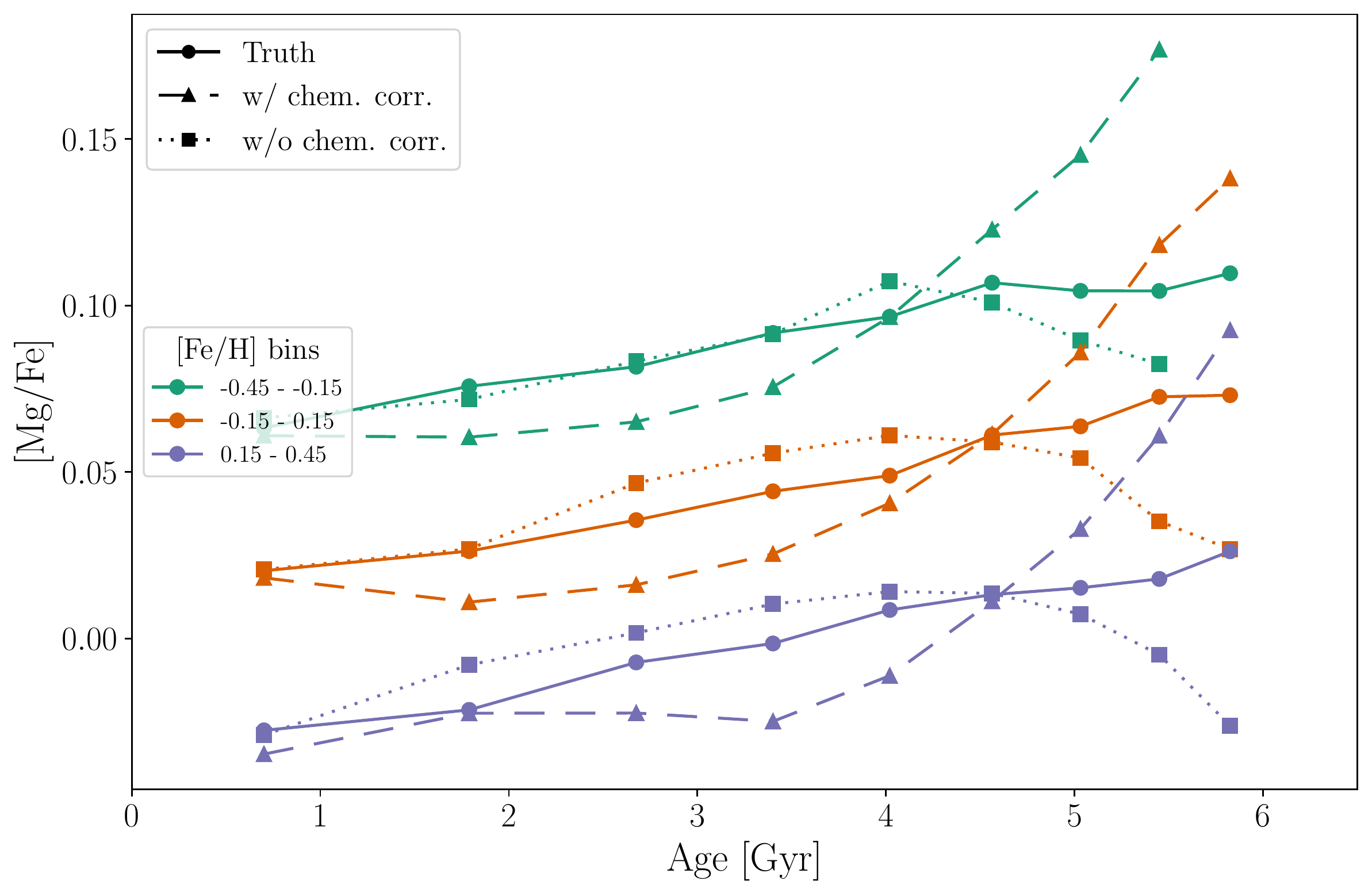}{0.48\textwidth}{(b)}
  }
  \caption{Same as Figure \ref{fig:trends_mw} but for the APOGEE-mocked red clump sample.}
  \label{fig:trends_apogee}
\end{figure*}

We now turn to the age-abundance trends for the RC sample.
Figure \ref{fig:trends_apogee} shows the age-abundance trends created from the PRFR models trained on the noisy RC sample.
On the left---where we take random samples from the conditional age distributions, we see that the predicted trends from the models trained on both the fiducial and chemically decorrelated datasets are in agreement with the true trends.
However, on the right panel, we observe interesting deviations from the truth when we take the modes of the conditional age distribution.
At large ages, the trends created from mode ages predicted from the model trained on the fiducial dataset (dashed line with triangles) have an upturn relative to the truth, while the trends from the mode ages predicted from the model trained on the chemically decorrelated dataset (dotted line with squares) have a downturn.

\begin{figure*}[htbp!]
  \centering
  \includegraphics[width=1\linewidth]{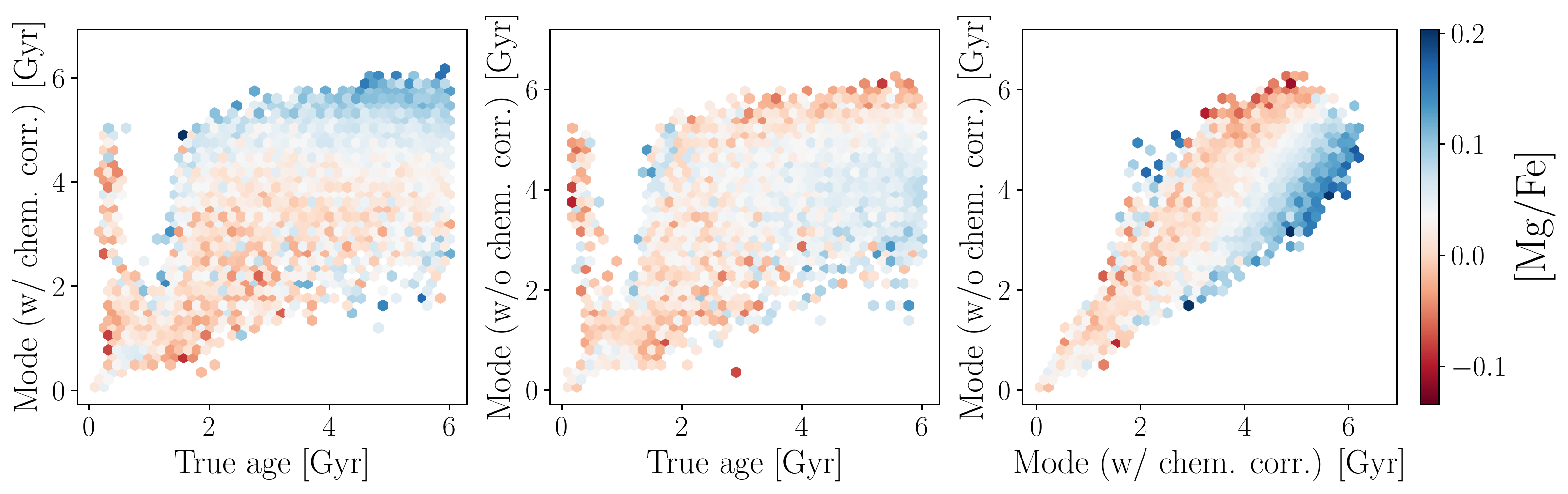}
  \caption{[Mg/Fe] dependence of age predictions. \textbf{Left:} True age plotted against the mode predicted age from the PRFR model trained on the fiducial dataset. \textbf{Center:} true age plotted against the mode predicted age from the model trained on the chemically-decorrelated dataset. \textbf{Right:} Mode predictions from models trained on the datasets with and without chemical correlations. Note the diagonal \mgfe\ gradient on the right, indicating \mgfe-dependence in the variation between the predictions.}
  \label{fig:mgfe-dep-age-mode-apogee}
\end{figure*}

The cause of this is that there is a \mgfe-dependent difference in the mode predictions between the fiducial and chemically decorrelated models.
In Figure \ref{fig:mgfe-dep-age-mode-apogee}, we explicitly show the structure of this \mgfe\ dependence.
For low \mgfe\ stars, the decorrelated predictions tend to be older than the fiducial predictions, and for high [Mg/Fe] stars, they tend to be younger.
Even though the two sets of mode age predictions have the same overall distributions, they differ for individual predictions, and that variation depends on \mgfe.
In particular, the right panel shows a diagonal gradient (i.e., across the 1:1 line) which indicates that there is \mgfe\ dependence in the variation between the two predictions.
So at, for example, low \mgfe\ values, one would expect to be in the red region of the plot where the prediction from the chemically decorrelated model is older than the prediction from the chemically correlated model.
Another way to think of this is that for a given star with some particular age, the \mgfe\ dependence of the model predictions means that the chemically correlated model is associated with a higher \mgfe\ value than the chemically decorrelated model.
Thus, there is an upturn in the trends for the chemically correlated lines and a downturn for the decorrelated lines.

We can also choose to use random samples from the posterior instead of modes.
Since random samples more representative of the full distribution of predictions, the variation of variation of \mgfe\ in true age and predicted age space is similar between the fiducial and chemically correlated cases, and thus the difference in the trends is much smaller than for modes.
As a result, using random samples instead of point estimates can lead to more faithful reconstructions of the age-abundance trends (as seen in Figure \ref{fig:trends_apogee}).

\needspace{2\baselineskip}
\section{Summary}\label{sec:summary}

In this paper, we have introduced a framework for estimating probabilistic stellar ages from stellar parameters.
We show that this framework is able to accurately and precisely recover ages, even in the face of observational noise, by applying it to the simple test case of a simulated dataset of Milky Way stars.
Using SHAP values, we interpret the model predictions and show how each variable drives age predictions, and how the importance of each of those variables varies across parameter space.
While in the fiducial case the PRFR model uses both stellar evolution and Galactic chemical correlations, we show that in a dataset where chemistry is explicitly decorrelated from ages, the only age information that is contained in \mgfe\ exists in the correlations it has with \feh\ (and by extension, stellar evolution).

Using these explanatory tools to compare the results before and after chemical decorrelation, we find that for the Milky Way sample, \logg\ and \teff\ are almost entirely responsible for driving predictions.
We then test our framework on a more difficult problem by applying it to a sample of red clump stars, where the range of \logg\ and \teff\ values are significantly restricted.
We show that in this case, the model is less reliant on stellar evolution, and instead proxies age information through chemical correlations.

Finally, we show that our framework is able to accurately reproduce age-abundance trends in the Milky Way sample.
However, for the red clump sample, our model uses information that is not intrinsic to the star (Galactic chemistry) to infer an intrinsic property (age).
As a result of this, use of ages predicted by the model can lead to misleadingly strong or weak age-abundance trends.

Our analysis here could be extended in many ways.
The most immediate extension would be to apply the framework to estimate stellar ages directly from stellar spectra, instead of from stellar labels.
Given that the number of features would increase significantly (depending on the resolution of the spectra), the calculation of exact SHAP values would no longer be feasible.
Thus, the use of approximate methods like Shapley sampling values or kernel SHAP \citep{Lundberg2017} and its extensions \citep[e.g.,][]{Aas2020} would be necessary, and may allow for the quantification of importance of, for example, different spectral lines.
Another extension would be to test models other than PRFR.
For example, likelihood-based models such as normalizing flows may allow for more rigorous statistical treatment of measurement errors, and the use of a generative, rather than discriminative, model provide more causal structure that could help to eliminate the bias introduced by the use of non-intrinsic information when inferring ages.
As SHAP values are model agnostic, they should be applicable in the same way to those new methods.

\section*{Acknowledgements}

We thank Neige Frankel for providing data and valuable input. Y.S.T. acknowledges financial support from the Australian Research Council through DECRA Fellowship DE220101520.

\software{
  NumPy \citep{Harris2020},
  Matplotlib \citep{Hunter2007},
  Seaborn \citep{Waskom2021},
  scikit-learn \citep{scikit-learn},
  cmasher \citep{cmasher}
}

\bibliography{sources-zotero, sources}

\appendix

\section{Bias correction and posterior calibration}\label{appendix:calibration}

\begin{figure}[htbp!]
  \centering
  \includegraphics[width=1\linewidth]{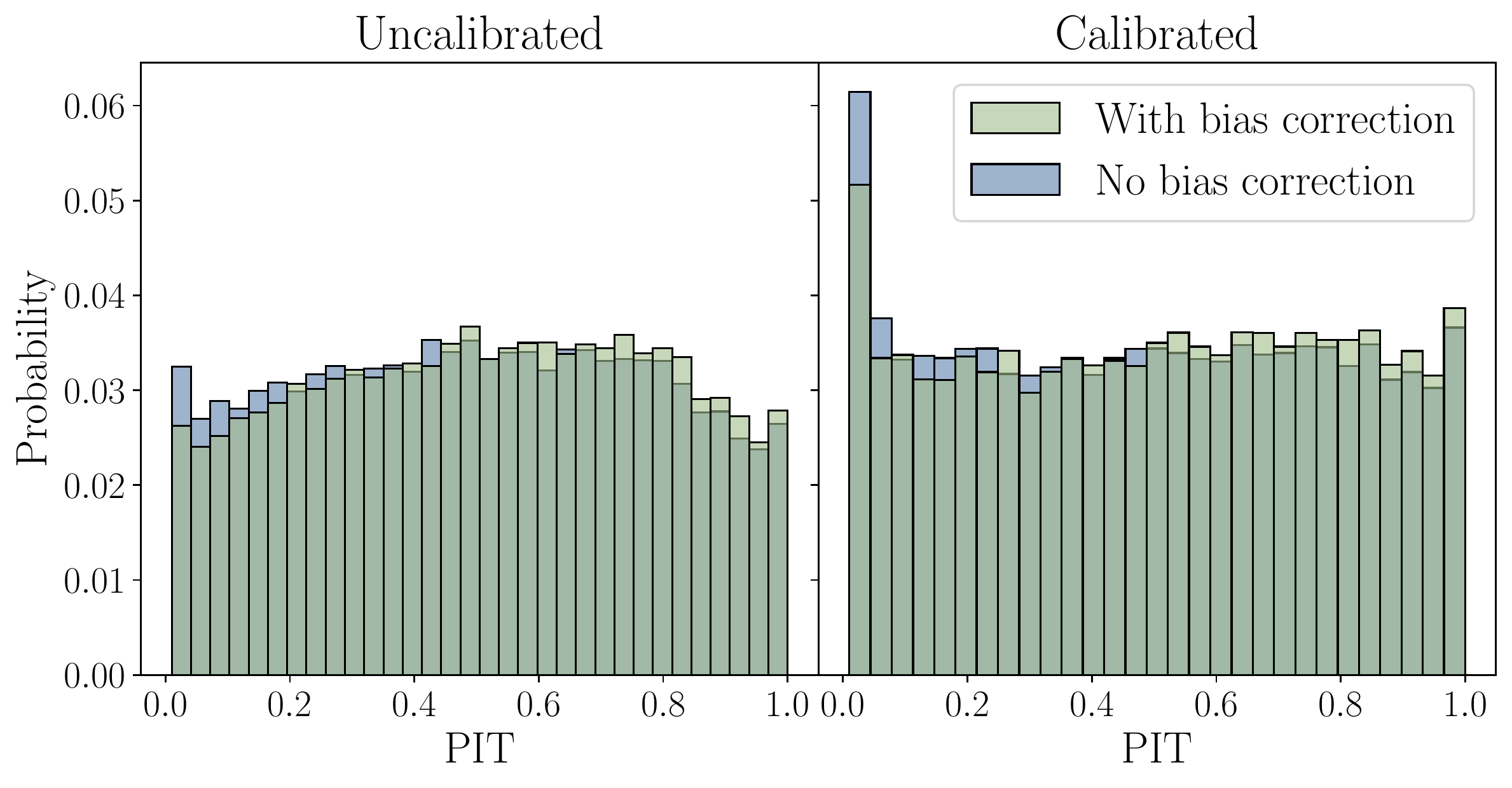}
  \caption{PIT histograms. The left panel shows the results from the uncalibrated models and the right panel shows the results from the calibrated models. In both panels, the blue histogram has no bias correction, and the green histogram has a bias correction applied. The calibration procedure makes the distribution of PIT values more uniform, and the bias correction generally shifts predictions to slightly lower masses.}
  \label{fig:pit-hist}
\end{figure}

We can assess the impact of the bias correction and posterior calibration by looking at PIT histograms (see Section \ref{sec:rf}).
After training bias correction and calibration models for our random forest on a validation set, we can calculate PIT values on a test set and plot the histograms.
Figure \ref{fig:pit-hist} shows the PIT histograms for calibrated and uncalibrated models with and without bias corrections for a small test set on the RC sample.

Moving from the left panel to the right panel, we see that the PIT histograms become more uniform, indicating that the posterior intervals are better calibrated.
Without calibration, the PIT histograms are too peaked in the center, indicating that the uncalibrated posteriors are too wide.
The cost of our calibration procedure is to occasionally underestimate the error on some stars, leading to slightly higher bin counts at the edges of the histograms.

The bias correction appears in the difference between the blue and the green histograms; its effect is small but noticeable.
In both the uncalibrated and calibrated models, the effect is to redistribute some of the counts in the left bins of the histogram (particularly the left-most bin) to the bins on the right.
In the calibrated case, the bias correction reduces the excess fraction at the left of the PIT histogram from roughly 3\% to roughly 2\%.
This indicates that overall, the bias correction serves to decrease the mass predictions, so that the true mass values which tended to be too small are now closer to the middle of the posteriors.
In fact, we find that the distribution of bias corrections on a test set is well described by a (location-scale) $t$ distribution with degrees of freedom $\nu=5.55$, location $\mu=-0.009$, and scale $\sigma=0.008$.

\end{CJK*}
\end{document}